\documentclass[11pt]{article}
\usepackage{amssymb,amsmath,amsfonts}
\usepackage{graphicx}
\usepackage{graphics}
\usepackage{eepic,epsfig}

\makeatletter
\@addtoreset{equation}{section}

\makeatother

\begin{document}

\title{Topological effects in the polarization of the Fulling-Rindler vacuum}
\author{A. A. Saharian$^{1}$,\, G. V. Mirzoyan$^{1}$, V. S. Torosyan$^{2}$ \\
\\
\textit{$^1$ Institute of Physics, Yerevan State University,}\\
\textit{1 Alex Manoogian Street, 0025 Yerevan, Armenia} \vspace{0.3cm}\\
\textit{$^{2}$ Institute of Mechanics RA,} \\
\textit{24b Marshall Baghramian Ave., 0019 Yerevan, Armenia}}
\maketitle

\begin{abstract}
We investigate how the compactification of some spatial dimensions in
Rindler spacetime affects the vacuum expectation values (VEVs) of the field
squared and the energy-momentum tensor for a charged scalar field prepared
in the Fulling-Rindler vacuum state. A toral compactification is considered
with quasi-periodicity conditions on the field operator along the compact
dimensions. The phases of these conditions are interpreted in terms of the
magnetic flux enclosed by the compact dimensions. For a general number of
spatial dimensions, the components of the VEVs are explicitly separated into
terms that correspond to the VEVs in the Minkowski vacuum and the terms
induced by the difference in the local characteristics of the
Fulling-Rindler and Minkowski vacua. As a limiting case, the VEVs are
retrieved in Rindler spacetime with trivial topology. We demonstrate that,
for non-zero phases in the periodicity conditions, the vacuum
energy-momentum tensor exhibits off-diagonal components with indices in the
compact subspace. Near the Rindler horizon, the leading terms in the
asymptotic expansions of the field squared and the diagonal components of
the energy-momentum tensor coincide with the corresponding VEVs in Rindler
spacetime with trivial topology. The off-diagonal components of the
energy-momentum tensor vanish on the Rindler horizon. For small
accelerations, the difference between the VEVs in the Fulling-Rindler and
Minkowski vacua is exponentially small. The exception to this is a massless
field with zero phases in the periodicity conditions. In this special case,
the difference decays according to a power law as a function of
acceleration. As applications, we consider the VEVs near the horizon of
cylindrical black holes and in curved backgrounds that are conformally
related to the Rindler spacetime.
\end{abstract}


\bigskip

\section{Introduction}

The canonical quantization of fields involves three main steps (see, e.g., 
\cite{Witt75,Birr82} for general curved background spacetimes). First, a
complete set of solutions to the classical field equations is selected. In
flat spacetime, plane waves are typically chosen, which describe states with
certain momenta. Next, the field operator is expanded over the selected
solutions, with the annihilation and creation operators being introduced as
expansion coefficients. The commutation relations between these operators
are obtained from the corresponding commutators for the field and
generalized momentum operators. The third step of quantization involves
constructing the Fock space of states, starting from the vacuum state. The
latter is defined as the state of the quantum field that is nullified by the
action of the annihilation operator. Particle states are constructed acting
by the creation operator on the vacuum state.

From the above, it follows that the notion of a vacuum is generally
observer-dependent and is determined by the choice of mode functions
employed in the quantization procedure \cite{Birr82,Full73}. Different
choices may lead to inequivalent vacuum states and, consequently, to
different predictions for local physical observables. A well-known example
is provided by the Minkowski and Fulling-Rindler (FR) vacua in flat
spacetime. While the Minkowski vacuum is natural for inertial observers, the
FR vacuum corresponds to the quantization of fields in the reference frame
of uniformly accelerated observers. The latter is closely related to the
Unruh effect, according to which uniformly accelerated observers perceive
the Minkowski vacuum as a thermal state \cite{Cris08}.

The study of quantum fields in the FR vacuum is motivated by several
reasons. By virtue of the equivalence principle, quantum fluctuations
perceived by accelerated observers are expected to share important features
with those occurring in gravitational backgrounds. In particular, the
Rindler geometry represents the leading approximation to the near-horizon
region of a broad class of black-hole spacetimes. Consequently,
investigations of vacuum effects in the FR vacuum provide valuable insight
into quantum phenomena in curved spacetime and black-hole physics. In
addition, the Rindler metric is conformally related to several physically
relevant geometries, including de Sitter spacetime and
Friedmann-Robertson-Walker cosmological models with negative spatial
curvature \cite{Birr82}. The vacuum expectation values (VEVs) for
conformally invariant fields in those backgrounds can be obtained from the
corresponding Rindler quantities by means of conformal transformations.
Furthermore, Rindler wedges in anti-de Sitter (AdS) spacetime play an
important role in holography and the AdS/CFT correspondence \cite{Empa99}-%
\cite{Zhan23}.

Another important aspect of quantum field theory concerns the influence of
nontrivial topology on vacuum fluctuations. The modification of the spectrum
of zero-point fluctuations by boundaries, compact dimensions, or external
fields gives rise to Casimir-type effects, which have been extensively
investigated in a variety of physical settings \cite{Most97}-\cite{Casi11}.
In spacetimes with compact spatial dimensions, the VEVs become sensitive to
the global properties of the background geometry and to the periodicity
conditions imposed on quantum fields. For charged fields, these effects
depend on magnetic fluxes enclosed by compact dimensions and provide vacuum
manifestations of the Aharonov-Bohm effect.

Vacuum polarization effects in the FR vacuum have been investigated in a
variety of geometrical configurations, including accelerated mirrors,
plates, branes, and Rindler-like spacetimes with nontrivial topology \cite%
{Cand77}-\cite{Saha26}. These studies have shown that acceleration, boundary
conditions, and spatial topology can significantly modify local vacuum
characteristics. In the present paper, we investigate vacuum polarization
effects for a charged massive scalar field in a $(D+1)$-dimensional Rindler
spacetime with spatial topology $R^{p}\times (S^{1})^{q}$. General
quasiperiodicity conditions are imposed along the compact dimensions. In the
presence of a constant gauge field, the corresponding phases are shifted by
the enclosed magnetic fluxes, leading to Aharonov-Bohm-type effects in the
VEVs. By employing a representation of the Hadamard function, we evaluate
the VEVs of the field squared and the energy-momentum tensor in the FR
vacuum and analyze their dependence on the proper acceleration,
compactification lengths, field mass, curvature coupling parameter, and
magnetic fluxes. The obtained results are further applied to the
near-horizon geometry of cylindrical black holes and topological black holes
with toroidal horizons \cite{Lemo95}-\cite{Gaet17}. Cylindrical black holes
in AdS background and related to them black strings appear naturally in
supergravity, string theory compactifications, AdS/CFT correspondence, and
braneworld models.

The paper is organized as follows. In the next section, we present the setup
of the problem and scalar field modes. The VEV of the field squared is
studied in \ Section \ref{sec:phi2}. Section \ref{sec:EMT} is devoted to the
investigation of the VEV of the energy-momentum tensor. Applications to
cylindrical configurations with toroidal dimensions and curved backgrounds
conformally related to the Rindler spacetime are discussed in Section \ref%
{sec:Cyl}. The main results are summarized in Section \ref{sec:Conc}. In
Appendix \ref{sec:Mink}, we describe the zeta function regularization for
the VEVs in the Minkowski spacetime with topology $R^{p+1}\times (S^{1})^{q}$
for a general case of phases in the periodicity conditions along toral
dimensions. Alternative representations of the VEVs, adapted for asymptotic
analysis, are considered in Appendix \ref{sec:Alt}.

\section{Scalar field modes in Rindler spacetime with compact dimensions}

\label{sec:Modes}

We consider a $(D+1)$-dimensional, locally Rindler spacetime covered by the
coordinates system $x^{i}=\left( x^{0}=\tau ,x^{1}=\xi ,\mathbf{x}\right) $
with dimensionless time $\tau \in (-\infty ,+\infty )$ and with $\xi \in
\lbrack 0,\infty )$. The $(D-1)$-dimensional flat subspace with coordinates $%
\mathbf{x}$ has a topology $R^{p}\times (S^{1})^{q}$, $p+q=D-1$ (for a
review of quantum field theory in spacetimes with toroidal topology see \cite%
{Khan14}). The sets of uncompact and compact coordinates will be denoted by $%
\mathbf{x}_{p}=(x^{2},\ldots ,x^{p+1})$ and $\mathbf{x}_{q}=(x^{p+2},\ldots
,x^{D})$, respectively. For these coordinates we have $-\infty <x^{i}<\infty 
$, $i=2,..,p+1$, and $0\leq x^{l}\leq L_{l}$, $l=l=p+2,...,D$. The line
element is expressed as 
\begin{equation}
ds^{2}=g_{il}dx^{i}dx^{l}=\xi ^{2}d\tau ^{2}-d\xi ^{2}-d\mathbf{x}^{2}.
\label{ds2}
\end{equation}%
The worldline with fixed coordinates $(\xi ,\mathbf{x})$ describes an
observer moving with constant proper acceleration $1/\xi $ with respect to
an inertial observer. For the corresponding proper time $t_{\mathrm{p}}$ one
has $t_{\mathrm{p}}=\xi \tau $. By the coordinate transformation%
\begin{equation}
t_{\mathrm{M}}=\xi \sinh \tau ,\;x_{\mathrm{M}}^{1}=\xi \cosh \tau ,\;%
\mathbf{x}_{\mathrm{M}}=\mathbf{x},  \label{txM}
\end{equation}%
the line element takes a locally Minkowskian form 
\begin{equation}
ds^{2}=\eta _{il}dx_{\mathrm{M}}^{i}dx_{\mathrm{M}}^{l}=dt_{\mathrm{M}%
}^{2}-\left( dx_{\mathrm{M}}^{1}\right) ^{2}-d\mathbf{x}_{\mathrm{M}}^{2}.
\label{ds2M}
\end{equation}%
In the geometry under consideration, this line element describes a locally
Minkowski spacetime with spatial topology $R^{p+1}\times (S^{1})^{q}$ and
with the lengths of compact dimensions $L_{l}$, $l=2,..,p+1$. The
coordinates $x^{i}$ in (\ref{ds2}) cover the Rindler wedge $x_{\mathrm{M}%
}^{1}>|t_{\mathrm{M}}|$ of the Minkowski spacetime. The value $\xi =0$
corresponds to the Rindler horizon.

We are interested in the effects of nontrivial spatial topology on physical
characteristics of the FR vacuum for a charged scalar field $\varphi (x)$
with mass $m$, charge $e$, and curvature coupling parameter $\zeta $. In the
geometry under consideration, the background spacetime is flat and the
coupling parameter does not appear in the field equation. However, the
expression for the metric energy-momentum tensor, and consequently the
corresponding VEV, depends on $\zeta $. In the presence of a classical gauge
field $A_{i}$, the field equation becomes%
\begin{equation}
\left( g^{il}D_{i}D_{l}+m^{2}\right) \varphi (x)=0,  \label{Feq}
\end{equation}%
where the gauge extended covariant derivative operator is given by $%
D_{l}=\nabla _{l}+ieA_{l}$, and $\nabla _{l}$ corresponds to the standard
covariant derivative associated with the metric tensor $g_{il}$ defined by (%
\ref{ds2}). To fix the dynamics, it is necessary to specify the periodicity
conditions in the compact subspace. We will impose the quasiperiodicity
condition%
\begin{equation}
\varphi (\tau ,\xi ,\mathbf{x}+L_{l}\mathbf{e}_{l})=e^{i\alpha _{l}}\varphi
(\tau ,\xi ,\mathbf{x}),  \label{PC}
\end{equation}%
along the compact dimension $x^{l}$, $l=p+2,\ldots ,D$, with constant phases 
$\alpha _{l}$ and with $\mathbf{e}_{l}$ being the unit vector along that
direction.

For a gauge field, the simplest configuration with constant components $%
A_{l} $ will be assumed. The components along uncompact dimensions are
removed from the problem by a gauge transformation. Due to non-trivial
topology, this is not the case for components in the compact subspace. The
corresponding gauge transformation affects the phases in the periodicity
conditions for gauge transformed scalar field. Indeed, let $A_{l}=\mathrm{%
const}$ be the component along direction $x^{l}$ with $l=p+2,\ldots ,D$.
Under the gauge transformation $\varphi ^{\prime }(x)=e^{ie\chi }\varphi (x)$%
, $A_{l}^{\prime }=A_{l}-\partial _{l}\chi $ with the function $\chi
=A_{l}x^{l}$, the vector potential in the new gauge becomes zero: $A_{\mu
}^{\prime }=0$. For the gauge transformed field, the periodicity condition
reads 
\begin{equation}
\varphi ^{\prime }(\tau ,\xi ,\mathbf{x}+L_{l}\mathbf{e}_{l})=e^{i\tilde{%
\alpha}_{l}}\varphi ^{\prime }(\tau ,\xi ,\mathbf{x}),\;\tilde{\alpha}%
_{l}=\alpha _{l}+eA_{l}L_{l}.  \label{PCn}
\end{equation}%
Therefore, although the gauge transformation removes the constant vector
potential from the field equation, it also changes the phases in the
periodicity conditions. As a result, the components along the compact
dimension are physically relevant. Particularly, the VEVs of physical
observables depend on these components. This is an Aharonov-Bohm-type
effect. Note that the product $A_{l}L_{l}$ in the expression of the new
phase $\tilde{\alpha}_{l}$ can be formally interpreted in terms of the
magnetic flux $\Phi _{l}=-A_{l}L_{l}$, enclosed by the compact dimension $%
x^{l}$. Introducing the flux quantum $\Phi _{0}=2\pi /e$, the phase is
expressed as $\tilde{\alpha}_{l}=\alpha _{l}-2\pi \Phi _{l}/\Phi _{0}$. The
discussion below will be given in terms of the gauge transformed field $%
\varphi ^{\prime }(x)$ omitting the prime.

The complete set of positive (upper sign) and negative (lower sign) energy
mode functions for the field $\varphi (x)$, obeying the equation (\ref{Feq})
with $A_{l}=0$ and specifying the FR vacuum state, are given by (see \cite%
{Kota22}) 
\begin{equation}
\varphi _{\omega ,\mathbf{k}}^{(\pm )}(x)=\frac{\sqrt{\sinh \left( \pi
\omega \right) }}{2^{\frac{p}{2}}\pi ^{\frac{p}{2}+1}\sqrt{V_{q}}}e^{i%
\mathbf{k}\cdot \mathbf{x}\mp i\omega \tau }K_{i\omega }(\lambda \xi
),\;\lambda =\sqrt{k^{2}+m^{2}},  \label{Modes}
\end{equation}%
where $\omega $ is the energy related to the proper time $\tau $, $\mathbf{k}%
=(k^{2},k^{3},\ldots ,k^{D})$ is the momentum in the subspace with
coordinates $\mathbf{x}$. In (\ref{Modes}), $V_{q}=L_{p+2}....L_{D}$ is the
volume of compact subspace, $K_{\nu }(z)$ is the Macdonald function, $k=|%
\mathbf{k}|$, and $\mathbf{k}\cdot \mathbf{x}=\sum_{l=2}^{D}k^{l}x^{l}$. For
the component of the momentum in the uncompact subspace we have $\mathbf{k}%
_{p}=(k^{2},\ldots ,k^{p+1})$ with $-\infty <k^{i}<+\infty $ for $i=2,\ldots
,p+1$. The components in the compact subspace, $\mathbf{k}%
_{q}=(k^{p+2},\ldots ,k^{D})$, are quantized by the periodicity conditions
and the corresponding eigenvalues become%
\begin{equation}
k_{l}=\left( 2\pi n_{l}+\tilde{\alpha}_{l}\right) /L_{l},\;n_{l}=0,\pm 1,\pm
2,\ldots .,  \label{kltild}
\end{equation}
for $l=p+2,\ldots ,D$. For the energy measured by an observer with fixed $%
(\xi ,\mathbf{x})$ one has $\omega _{\mathrm{p}}=\omega /\xi $ and in the
exponent of (\ref{Modes}) we can write $\omega \tau =\omega _{\mathrm{p}}t_{%
\mathrm{p}}$.

\section{VEV of the field squared}

\label{sec:phi2}

Having the mode functions, we can evaluate the Hadamard function, defined as
the VEV $G(x,x^{\prime })=\left\langle 0\right\vert \varphi (x)\varphi
^{\dagger }(x^{\prime })+\varphi ^{\dagger }(x^{\prime })\varphi
(x)\left\vert 0\right\rangle $, by using the corresponding mode sum formula.
Here, $\left\vert 0\right\rangle $ stands for the FR vacuum state. The
latter differs from the Minkowski vacuum for which we use the notation $%
\left\vert 0\right\rangle _{\mathrm{M}}$. In \cite{Kota22}, the following
representation is derived for the Hadamard function: 
\begin{eqnarray}
G(x,x^{\prime }) &=&G_{\mathrm{M}}(x,x^{\prime })-2\sum_{\mathbf{n}_{q}}%
\frac{\omega _{\mathbf{n}_{q}}^{p}e^{i{\mathbf{k}}_{q}\cdot \Delta {\mathbf{x%
}}_{q}}}{\left( 2\pi \right) ^{\frac{p}{2}+1}V_{q}}\int_{0}^{\infty
}dT\sum_{s=\pm 1}\frac{1}{\left( T-s\Delta \tau \right) ^{2}+\pi ^{2}} 
\notag \\
&&\times f_{\frac{p}{2}}\left( \omega _{\mathbf{n}_{q}}\sqrt{|\Delta {%
\mathbf{x}}_{p}|^{2}+\xi ^{2}+\xi ^{\prime 2}+2\xi \xi ^{\prime }\cosh T}%
\right) ,  \label{Had}
\end{eqnarray}%
where $\Delta {\mathbf{x}}_{q}={\mathbf{x}}_{q}-{\mathbf{x}_{q}^{\prime }}$, 
$\Delta {\mathbf{x}}_{p}={\mathbf{x}}_{p}-{\mathbf{x}_{p}^{\prime }}$, and
we have introduced the notations 
\begin{equation}
\omega _{\mathbf{n}_{q}}=\sqrt{\mathbf{k}_{q}^{2}+m^{2}},\;f_{\nu
}(z)=z^{-\nu }K_{\nu }(z).  \label{fnu}
\end{equation}%
Here and below, $\mathbf{n}_{q}=(n_{p+2},n_{p+3},\ldots ,n_{D})$ and $\sum_{%
\mathbf{n}_{q}}=\prod_{l=p+2}^{D}\sum_{n_{l}=-\infty }^{+\infty }$. In (\ref%
{Had}), $G_{\mathrm{M}}(x,x^{\prime })$ is the Hadamard function in a
locally Minkowskian spacetime with the spatial topology $R^{p+1}\times
(S^{1})^{q}$ and with coordinates $(x^{p+2},\ldots ,x^{D})$ compactified to
a torus $(S^{1})^{q}$ with the same lengths of the compact subspace as in
the Rindler problem. Note that the divergences for $G(x,x^{\prime })$ and $%
G_{\mathrm{M}}(x,x^{\prime })$ in the coincidence limit $x^{\prime
}\rightarrow x$ are the same and the last term in (\ref{Had}) is finite in
that limit (except the points on the horizon).

The VEV of the field squared, defined as $\left\langle \varphi \varphi
^{\dagger }\right\rangle =\left\langle 0\right\vert \varphi (x)\varphi
^{\dagger }(x)+\varphi ^{\dagger }(x)\varphi (x)\left\vert 0\right\rangle /2$%
, is evaluated by using the formula%
\begin{equation}
\langle \varphi \varphi ^{\dagger }\rangle =\frac{1}{2}\lim_{x^{\prime
}\rightarrow x}G(x,x^{\prime }).  \label{phi2lim}
\end{equation}%
By making use the expression (\ref{Had}) of the Hadamard function, we get%
\begin{equation}
\langle \varphi \varphi ^{\dagger }\rangle =\langle \varphi \varphi
^{\dagger }\rangle _{\mathrm{M}}-\frac{1}{\left( 2\pi \right) ^{\frac{p}{2}%
+1}V_{q}}\sum_{\mathbf{n}_{q}}\omega _{\mathbf{n}_{q}}^{p}\int_{0}^{\infty
}dy\,\frac{f_{p/2}\left( 2\xi \omega _{\mathbf{n}_{q}}\cosh y\right) }{%
y^{2}+\pi ^{2}/4}.  \label{phi2}
\end{equation}%
where 
\begin{equation}
\langle \varphi \varphi ^{\dagger }\rangle _{\mathrm{M}}=\frac{m^{D-1}}{%
(2\pi )^{\frac{D+1}{2}}}\sideset{}{'}{\sum}_{\mathbf{n}_{q}}\cos (\mathbf{n}%
_{q}\cdot \boldsymbol{\tilde{\alpha}})f_{\frac{D-1}{2}}(mg_{\mathbf{n}_{q}})
\label{phi2M}
\end{equation}%
is the VEV for the Minkowski vacuum (see Appendix \ref{sec:Mink}). In (\ref%
{phi2M}), $\boldsymbol{\tilde{\alpha}}=(\tilde{\alpha}_{p+2},\ldots ,\tilde{%
\alpha}_{D})$, $\mathbf{n}_{q}\cdot \boldsymbol{\tilde{\alpha}}%
=\sum\nolimits_{i=p+2}^{D}n_{i}\tilde{\alpha}_{i}$, and%
\begin{equation}
g_{\mathbf{n}_{q}}=\sqrt{\sum\nolimits_{i=p+2}^{D}n_{i}^{2}L_{i}^{2}}.
\label{geL}
\end{equation}%
The prime in (\ref{phi2M}) means that the term $\mathbf{n}_{q}=\mathbf{0}%
_{q}=(0,\ldots ,0)$ should be excluded from the sum. In the special case $%
\tilde{\alpha}_{i}=0$, $i=p+2,\ldots ,D$, and for a massless field we have $%
\omega _{\mathbf{n}_{q}}=0$ for $\mathbf{n}_{q}=\mathbf{0}_{q}$. In this
case, the contribution of the term $\mathbf{n}_{q}=\mathbf{0}_{q}$ in (\ref%
{phi2}) is obtained by the limiting transition $\omega _{\mathbf{n}%
_{q}}\rightarrow 0$ and is expressed as 
\begin{equation}
-\frac{\Gamma \left( \frac{p}{2}\right) \xi ^{-p}}{\left( 4\pi \right) ^{%
\frac{p}{2}+1}V_{q}}\int_{0}^{\infty }dy\,\frac{\cosh ^{-p}y}{y^{2}+\pi
^{2}/4}.  \label{ZeroContr}
\end{equation}%
Note that the Minkowskian part (\ref{phi2M}) is homogeneous. Both the
contributions in (\ref{phi2}) are even periodic functions of $\tilde{\alpha}%
_{l}$ with the period $2\pi $. In terms of the magnetic flux, this
corresponds to the periodicity with the period of flux quantum $\Phi _{0}$.
The second term in the right-hand side of (\ref{phi2}) presents the
difference in the VEVs corresponding to the FR and Minkowski vacua. This
term is always negative and tends to zero in the limit $\xi \rightarrow 0$.

An alternative expression for the VEV $\langle \varphi \varphi ^{\dagger
}\rangle $ in the FR vacuum is obtained by applying to the series over $%
\mathbf{n}_{q}$ in (\ref{phi2}) the formula (\ref{SumTrans}) from Appendix %
\ref{sec:Alt} with $n=0$ and $b=2\xi \cosh y$. This gives 
\begin{equation}
\langle \varphi \varphi ^{\dagger }\rangle =\langle \varphi \varphi
^{\dagger }\rangle _{\mathrm{M}}-\frac{m^{D-1}}{(2\pi )^{\frac{D+1}{2}}}%
\sum_{\mathbf{n}_{q}}\cos (\mathbf{n}_{q}\cdot \boldsymbol{\tilde{\alpha}}%
)\int_{0}^{\infty }dy\,\frac{f_{\frac{D-1}{2}}(mw_{\mathbf{n}_{q}})}{%
y^{2}+\pi ^{2}/4},  \label{phi22}
\end{equation}%
with the notation%
\begin{equation}
w_{\mathbf{n}_{q}}=\sqrt{4\xi ^{2}\cosh ^{2}y+g_{\mathbf{n}_{q}}^{2}}.
\label{wnq}
\end{equation}%
Note that in the formulas (\ref{phi2M}) and (\ref{phi22}) we can make the
replacement 
\begin{equation}
\cos \left( \mathbf{n}_{q}\cdot \boldsymbol{\tilde{\alpha}}\right)
\rightarrow \prod\limits_{l=p+2}^{D}\cos (n_{l}\tilde{\alpha}_{l}).
\label{Repl}
\end{equation}%
In the limit $L_{i}\rightarrow \infty $, the only nonzero contribution in
the last term of (\ref{phi22}) comes from the term with $\mathbf{n}_{q}=%
\mathbf{0}_{q}$ and we get the result for the FR vacuum in the model with
trivial spatial topology \cite{Saha02}:%
\begin{equation}
\langle \varphi \varphi ^{\dagger }\rangle _{0}=-\frac{m^{D-1}}{(2\pi )^{%
\frac{D+1}{2}}}\int_{0}^{\infty }dy\,\frac{f_{\frac{D-1}{2}}(2m\xi \cosh y)}{%
y^{2}+\pi ^{2}/4}.  \label{phi2FR}
\end{equation}%
Note that in the Minkowski spacetime with trivial topology the VEV $\langle
\varphi \varphi ^{+}\rangle _{\mathrm{M}}$ is renormalized to zero. For a
massless field, the formula (\ref{phi2FR}) is reduced to%
\begin{equation}
\langle \varphi \varphi ^{\dagger }\rangle _{0}=-\frac{\Gamma \left( \frac{%
D-1}{2}\right) }{\left( 4\pi \right) ^{\frac{D+1}{2}}\xi ^{D-1}}%
\int_{0}^{\infty }dy\,\frac{\cosh ^{1-D}y}{y^{2}+\pi ^{2}/4}.
\label{phi2FRm0}
\end{equation}%
Comparing with (\ref{ZeroContr}), we see that for $\tilde{\alpha}_{i}=0$, $%
i=p+2,\ldots ,D$, and for a massless field the contribution of the term $%
\mathbf{n}_{q}=\mathbf{0}_{q}$ in (\ref{phi2}) is presented as $\langle
\varphi \varphi ^{\dagger }\rangle _{\mathrm{FR}}^{(p+2)}/V_{q}$, where $%
\langle \varphi \varphi ^{\dagger }\rangle _{\mathrm{FR}}^{(p+2)}$ is the
VEV in the $(p+2)$-dimensional Rindler spacetime with trivial topology.

Note that the term $\mathbf{n}_{q}=\mathbf{0}_{q}$ in the formula (\ref%
{phi22}) coincides with $\langle \varphi \varphi ^{+}\rangle _{0}$. Then, by
taking into account that $\int_{0}^{\infty }dy/(y^{2}+\pi ^{2}/4)=1$, we see
that the value at $\xi =0$ of the remaining part including the series with $%
\mathbf{n}_{q}\neq \mathbf{0}_{q}$ gives $-\langle \varphi \varphi ^{\dagger
}\rangle _{\mathrm{M}}$. From here, the decomposition%
\begin{equation}
\langle \varphi \varphi ^{\dagger }\rangle =\langle \varphi \varphi
^{\dagger }\rangle _{0}+\langle \varphi \varphi ^{\dagger }\rangle _{\mathrm{%
t}},  \label{phi23}
\end{equation}%
is obtained, where%
\begin{equation}
\langle \varphi \varphi ^{\dagger }\rangle _{\mathrm{t}}=-\frac{m^{D-1}}{%
(2\pi )^{\frac{D+1}{2}}}\sideset{}{'}{\sum}_{\mathbf{n}_{q}}\cos (\mathbf{n}%
_{q}\cdot \boldsymbol{\tilde{\alpha}})\int_{0}^{\infty }dy\,\frac{f_{\frac{%
D-1}{2}}(mw_{\mathbf{n}_{q}})-f_{\frac{D-1}{2}}(mg_{\mathbf{n}_{q}})}{%
y^{2}+\pi ^{2}/4}.  \label{phi2t}
\end{equation}%
The part $\langle \varphi \varphi ^{\dagger }\rangle _{\mathrm{t}}$ in the
total VEV is induced by nontrivial topology of the Rindler spacetime. This
topological contribution vanishes on the Rindler horizon.

The general expression (\ref{phi22}) is further simplified for a massless
field. By taking into account that%
\begin{equation}
f_{\nu }(x)\approx 2^{\nu -1}\Gamma (\nu )x^{-2\nu },\;x\ll 1,
\label{fsmall}
\end{equation}%
we get%
\begin{equation}
\langle \varphi \varphi ^{\dagger }\rangle =\langle \varphi \varphi
^{\dagger }\rangle _{\mathrm{M}}-\frac{\Gamma \left( \frac{D-1}{2}\right) }{%
4\pi ^{\frac{D+1}{2}}}\sum_{\mathbf{n}_{q}}\cos (\mathbf{n}_{q}\cdot 
\boldsymbol{\tilde{\alpha}})\int_{0}^{\infty }dy\,\frac{w_{\mathbf{n}%
_{q}}^{1-D}}{y^{2}+\pi ^{2}/4}.  \label{phi2m0}
\end{equation}%
We can also make the replacement (\ref{Repl}). The corresponding formula for
the Minkowskian part is obtained from (\ref{phi2M}):%
\begin{equation}
\langle \varphi \varphi ^{\dagger }\rangle _{\mathrm{M}}=\frac{\Gamma \left( 
\frac{D-1}{2}\right) }{4\pi ^{\frac{D+1}{2}}}\sideset{}{'}{\sum}_{\mathbf{n}%
_{q}}\frac{\cos (\mathbf{n}_{q}\cdot \boldsymbol{\tilde{\alpha}})}{g_{%
\mathbf{n}_{q}}^{D-1}},  \label{phi2m0M}
\end{equation}

Let us consider the behavior of the field squared in the asymptotic regions
of $\xi $. For small values of $\xi $ it is convenient to use the
representation (\ref{phi22}). From the series over $\mathbf{n}_{q}$ in the
last term we separate the contribution with $\mathbf{n}_{q}=\mathbf{0}_{q}$.
In the remaining part, in the leading order, we can directly put $\xi =0$.
With this substitution the integral over $y$ gives 1 and this part is
reduced to $\langle \varphi \varphi ^{\dagger }\rangle _{\mathrm{M}}$. This
cancels the Minkowskian part (the first term in the right-hand side of (\ref%
{phi22})). The dominant contribution to the VEV $\langle \varphi \varphi
^{+}\rangle $ comes from the term with $\mathbf{n}_{q}=\mathbf{0}_{q}$ which
coincides with (\ref{phi2FR}). Assuming $m\xi \ll 1$, the leading term in
the expansion near the horizon becomes%
\begin{equation}
\langle \varphi \varphi ^{\dagger }\rangle \approx -\frac{\Gamma \left( 
\frac{D-1}{2}\right) }{\left( 4\pi \right) ^{\frac{D+1}{2}}\xi ^{D-1}}%
\int_{0}^{\infty }dx\,\frac{\cosh ^{1-D}x}{x^{2}+\pi ^{2}/4}.
\label{phi2hor}
\end{equation}%
Comparing with (\ref{phi2FRm0}), we see that the leading term coincides with
the VEV for a massless scalar field in the Rindler spacetime with trivial
topology. This shows that near the horizon the topological contribution in
the VEV is small.

In the opposite limit $\xi \gg L_{i}$, it is more convenient to use the
representation (\ref{phi2}). By using the asymptotic expression for the
modified Bessel function for large arguments \cite{Abra}, we see that the
dominant contribution to the series over $\mathbf{n}_{q}$ comes from the
term with the smallest value of $\omega _{\mathbf{n}_{q}}$ that will be
denoted here by $\omega _{0}$. Assuming that $|\tilde{\alpha}_{i}|\leq \pi $
and $\omega _{0}\neq 0$, one has%
\begin{equation}
\omega _{0}=\sqrt{\sum\nolimits_{i=p+2}^{D}\tilde{\alpha}%
_{i}^{2}/L_{i}^{2}+m^{2}}.  \label{om0n}
\end{equation}%
In addition, the main contribution to the integral over $x$ comes from the
integration range near the lower limit. In this way we can see that, to the
leading order,%
\begin{equation}
\langle \varphi \varphi ^{\dagger }\rangle \approx \langle \varphi \varphi
^{\dagger }\rangle _{\mathrm{M}}-\frac{\omega _{0}^{p/2-1}e^{-2\xi \omega
_{0}}}{2^{p+1}\pi ^{p/2+2}V_{q}\xi ^{p/2+1}},  \label{phi2large}
\end{equation}%
and the difference of the VEVs in the Minkowski and FR vacua is
exponentially suppressed. For a massless field and for $\tilde{\alpha}_{i}=0$
we have $\omega _{0}=0$. In this case and for $\xi \gg L_{i}$, the dominant
contribution in (\ref{phi2}) comes from the term with $\omega _{\mathbf{n}%
_{q}}=0$ ($\mathbf{n}_{q}=\mathbf{0}_{q}$). This contribution is given by (%
\ref{ZeroContr}) and in the limit under consideration we get%
\begin{equation}
\langle \varphi \varphi ^{\dagger }\rangle \approx \langle \varphi \varphi
^{\dagger }\rangle _{\mathrm{M}}-\frac{\Gamma \left( \frac{p}{2}\right) \xi
^{-p}}{\left( 4\pi \right) ^{\frac{p}{2}+1}V_{q}}\int_{0}^{\infty }dy\,\frac{%
\cosh ^{-p}y}{y^{2}+\pi ^{2}/4}.  \label{phi2largem0}
\end{equation}%
Therefore, for a massless field with $\tilde{\alpha}_{i}=0$ we have power
law decay for large $\xi $ instead of exponential suppression in (\ref%
{phi2large}). Note that the asymptotic behavior (\ref{phi2largem0}) could
also be obtained from (\ref{phi2m0}), by taking into account that for $\xi
\gg L_{i}$ the dominant contribution comes from large values of $n_{i}$ and
we can replace the summation over $\mathbf{n}_{q}$ by integration.

The formulas (\ref{phi2large}) and (\ref{phi2largem0}) also describe the
behavior of the VEV for small values of the lengths of compact dimensions,
corresponding to $mL_{i}\ll 1$ and $L_{i}\ll \xi $. In this case, the
Minkowskian part is approximated by (\ref{phi2Mm0}) and it dominates in the
total VEV. In the opposite limit of large values of the lengths of compact
dimensions, $L_{i}\gg \xi ,1/m$, the VEV (\ref{phi2}) is dominated by the
last term and is approximated by $\langle \varphi \varphi ^{\dagger }\rangle
_{0}$.

The numerical results below will be given for the geometry with a single
compact dimension $x^{D}$ of the length $L$. In this special case, one has $%
q=1$, $p=D-2$ and the general formulas are simplified to 
\begin{equation}
\langle \varphi \varphi ^{\dagger }\rangle =\langle \varphi \varphi
^{\dagger }\rangle _{\mathrm{M}}-\frac{1}{\left( 2\pi \right) ^{\frac{D}{2}}L%
}\sum_{n=-\infty }^{+\infty }\omega _{D}^{D-2}\int_{0}^{\infty }dx\,\frac{%
f_{D/2-1}\left( 2\xi \omega _{D}\cosh y\right) }{y^{2}+\pi ^{2}/4}.
\label{phi2q1}
\end{equation}%
with $k_{D}=\left( 2\pi n+\tilde{\alpha}_{D}\right) /L$ and $\omega _{D}=%
\sqrt{k_{D}^{2}+m^{2}}$. The Minkowskian part takes the form%
\begin{equation}
\langle \varphi \varphi ^{\dagger }\rangle _{\mathrm{M}}=\frac{2m^{D-1}}{%
(2\pi )^{\frac{D+1}{2}}}\sum_{n=1}^{\infty }\cos (n\tilde{\alpha}_{D})f_{%
\frac{D-1}{2}}\left( nmL\right) .  \label{phi2q1M}
\end{equation}%
An equivalent representation is obtained from (\ref{phi22}):%
\begin{equation}
\langle \varphi \varphi ^{\dagger }\rangle =\langle \varphi \varphi
^{\dagger }\rangle _{\mathrm{M}}-\frac{2m^{D-1}}{(2\pi )^{\frac{D+1}{2}}}%
\sum_{n=0}^{\infty }\delta _{n}\cos (n\tilde{\alpha}_{D})\int_{0}^{\infty
}dx\,\frac{f_{\frac{D-1}{2}}(m\sqrt{4\xi ^{2}\cosh ^{2}y+n^{2}L^{2}})}{%
y^{2}+\pi ^{2}/4},  \label{phi2q1b}
\end{equation}%
where $\delta _{0}=1/2$ and $\delta _{n}=1$ for $n\neq 0$. For a massless
field, this is reduced to%
\begin{equation}
\langle \varphi \varphi ^{\dagger }\rangle =\langle \varphi \varphi
^{\dagger }\rangle _{\mathrm{M}}-\frac{\Gamma \left( \frac{D-1}{2}\right) }{%
2\pi ^{\frac{D+1}{2}}}\sum_{n=0}^{\infty }\delta _{n}\cos (n\tilde{\alpha}%
_{D})\int_{0}^{\infty }dx\,\frac{\left( 4\xi ^{2}\cosh
^{2}y+n^{2}L^{2}\right) ^{\frac{1-D}{2}}}{y^{2}+\pi ^{2}/4}.
\label{phi2q1m0}
\end{equation}%
with the Minkowskian VEV%
\begin{equation}
\langle \varphi \varphi ^{\dagger }\rangle _{\mathrm{M}}=\frac{\Gamma \left( 
\frac{D-1}{2}\right) }{2\pi ^{\frac{D+1}{2}}L^{D-1}}\sum_{n=1}^{\infty }%
\frac{\cos (n\tilde{\alpha}_{D})}{n^{D-1}}.  \label{phi2Mm0}
\end{equation}

\begin{figure}[tbph]
\begin{center}
\begin{tabular}{cc}
\epsfig{figure=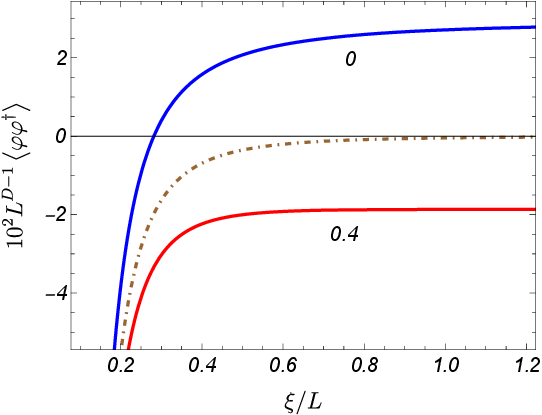,width=8cm,height=6.5cm} & \quad %
\epsfig{figure=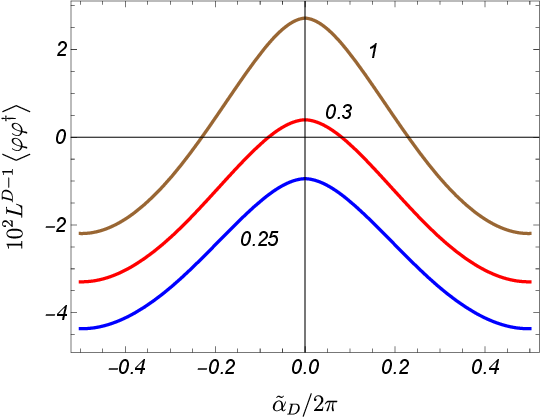,width=8cm,height=6.5cm}%
\end{tabular}%
\end{center}
\caption{The VEV of the field squared for a massless field as a function of $%
\protect\xi /L$ in the $D=4$ model (left panel). The graphs are plotted for $%
\tilde{\protect\alpha}_{D}=0$ and $\tilde{\protect\alpha}_{D}/2\protect\pi %
=0.4$ (the numbers near the curves). The dot-dashed curve corresponds to the
VEV in the Rindler spacetime with trivial topology. The right panel presents
the dependence on $\tilde{\protect\alpha}_{D}/2\protect\pi $. The numbers
near the curves are the values of $\protect\xi /L$.}
\label{fig1}
\end{figure}

For a massless field the combination $L^{D-1}\langle \varphi \varphi
^{\dagger }\rangle $ is a function of the ratio $\xi /L$. On the left panel
of Fig. \ref{fig1} this combination is plotted versus $\xi /L$ for $D=4$.
The numbers near the curves correspond to the values of $\tilde{\alpha}%
_{D}/2\pi $. The dot-dashed line corresponds to the combination $%
L^{D-1}\langle \varphi \varphi ^{\dagger }\rangle _{0}$. The right panel of
Fig. \ref{fig1} displays the dependence of the VEV $\langle \varphi \varphi
^{\dagger }\rangle $ as a function of the ratio $\tilde{\alpha}_{D}/2\pi $
for a massless field and for different values of $\xi /L$ (numbers near the
curves).

For small accelerations we have $\xi /L\gg 1$ and the VEV $\langle \varphi
\varphi ^{\dagger }\rangle $ tends to the corresponding value in the $(4+1)$%
-dimensional Minkowski spacetime with a compact dimension $x^{4}$ (see (\ref%
{phi2Mm0})). For large accelerations, corresponding to $\xi /L\ll 1$ (points
near the Rindler horizon), the contribution coming from the part $\langle
\varphi \varphi ^{\dagger }\rangle _{0}$ (given by (\ref{phi2FRm0}))
dominates in the VEV.

\begin{figure}[tbph]
\begin{center}
\begin{tabular}{cc}
\epsfig{figure=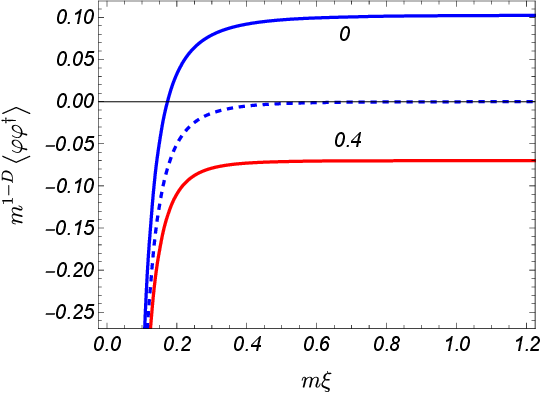,width=8cm,height=6.5cm} & \quad %
\epsfig{figure=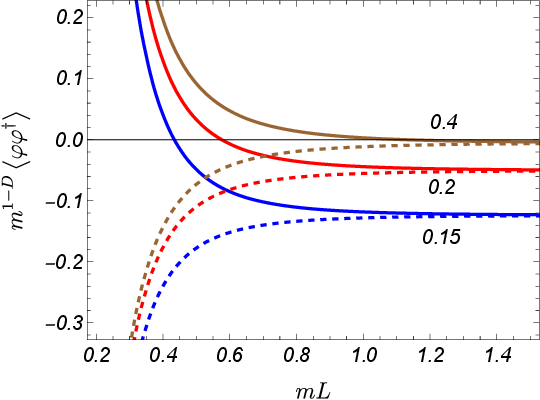,width=8cm,height=6.5cm}%
\end{tabular}%
\end{center}
\caption{The VEV of the field squared for a massive field as a function of
inverse proper acceleration (left panel) and the length of compact dimension
(right panel) in units of the Compton wavelength. The numbers near the
curves on the left and right panels are the values of $\tilde{\protect\alpha}%
_{D}/2\protect\pi $ and $m\protect\xi $, respectively. The full and dashed
curves on the right panel are plotted for $\tilde{\protect\alpha}_{D}/2%
\protect\pi =0$ and $\tilde{\protect\alpha}_{D}/2\protect\pi =0.4$.}
\label{fig2}
\end{figure}

Figure \ref{fig2} presents the VEV of the field squared for a massive field
as a function of the coordinate $\xi $ and length of the compact dimension $%
L $, in units of $1/m$. The left panel is plotted for $mL=0.5$ and the
numbers near the curves are the values of $\tilde{\alpha}_{D}/2\pi $. The
dot-dashed curve on the left panel presents the VEV in the Rindler spacetime
without compact dimensions. The numbers near the curves on the right panel
are the values of the $m\xi $ and the full and dashed curves correspond to
the values of $\tilde{\alpha}_{D}/2\pi =0$ and $\tilde{\alpha}_{D}/2\pi =0.4$%
, respectively. Similar to the case of a massless field, for large values of 
$m\xi $ (small acceleration) the VEV $\langle \varphi \varphi ^{\dagger
}\rangle $ is approximated by the Minkowskian VEV (\ref{phi2Mm0}). In the
opposite limit of large acceleration, the VEV tends to the Rinlder VEV in
the geometry with decompactified coordinate $x^{4}$. For small values of the
length of the compact dimension, $mL\ll 1$, the effects of acceleration are
small and the VEV $\langle \varphi \varphi ^{\dagger }\rangle $ is
approximated by the expectation value in the Minkowski spacetime with a
single compact dimension. In the region $mL\gg 1$ and for fixed $m\xi $, the
effects of nontrivial topology are small and the VEV tends to $\langle
\varphi \varphi ^{\dagger }\rangle _{0}$, given by (\ref{phi2FR}).

The above analysis showed that the VEV of the field squared in the
Fulling-Rindler vacuum may change sign depending on the acceleration. This
can lead to instabilities in theories of interacting fields. Consider, for
instance, a self-acting scalar field that interacts with a Dirac field $\psi 
$. In the case of quartic self-interaction, an example of the interaction
Lagrangian density is $\mathcal{L}_{\mathrm{int}}=-\lambda \left( \varphi
\varphi ^{\dagger }\right) ^{4}-g\varphi \varphi ^{\dagger }\bar{\psi}\psi $
with $\bar{\psi}=\psi ^{\dagger }\gamma ^{(0)}$ being the Dirac conjugate.
The appearance of the nonzero VEV $\langle \varphi \varphi ^{\dagger
}\rangle $ will contribute to the effective mass squared for both scalar and
Dirac fields. Depending on the sign of $\langle \varphi \varphi ^{\dagger
}\rangle $, the effective mass squared may become negative. This change in
sign can indicate an instability signaling a phase transition.

\section{Mean energy-momentum tensor}

\label{sec:EMT}

For the VEV\ of the energy-momentum tensor we use the formula%
\begin{equation}
\langle T_{ik}\rangle =\frac{1}{4}\lim_{x^{\prime }\rightarrow x}\left(
\partial _{i}\partial _{k}^{\prime }+\partial _{k}\partial _{i}^{\prime
}\right) G(x,x^{\prime })+\left[ \left( \zeta -1/4\right) g_{ik}\nabla
_{l}\nabla ^{l}-\zeta \nabla _{i}\nabla _{k}\right] \langle \varphi \varphi
^{\dagger }\rangle ,  \label{vevEMT}
\end{equation}%
where $\zeta $ is the curvature coupling parameter. The most important
special cases correspond to minimal and conformal couplings with $\zeta =0$
and $\zeta =\zeta _{D}\equiv (D-1)/4D$, respectively. Similar to the case of
the field squared, the VEV (\ref{vevEMT}) is decomposed into the Minkowskian
part and an additional term induced by the difference of the properties of
the Minkowski and FR vacua. The latter is finite and is directly obtained
from the corresponding contributions in the Hadamard function and the VEV of
the field squared.

The term containing the d'Alembertian operator, which is common to all
components, has the following form:%
\begin{equation}
\nabla _{l}\nabla ^{l}\langle \varphi \varphi ^{\dagger }\rangle =\frac{2^{1-%
\frac{p}{2}}}{\pi ^{\frac{p}{2}+1}V_{q}}\sum_{\mathbf{n}_{q}}\omega _{%
\mathbf{n}_{q}}^{p+2}\int_{0}^{\infty }dy\,\cosh ^{2}y\frac{pf_{p/2+1}\left(
u\right) +f_{p/2}\left( u\right) }{\pi ^{2}/4+y^{2}},  \label{Dalphi}
\end{equation}%
with the notation%
\begin{equation}
u=u(y)=2\xi \omega _{\mathbf{n}_{q}}\cosh y.  \label{u}
\end{equation}%
Here, we have used the relation%
\begin{equation}
u^{2}f_{\nu +1}\left( u\right) =2\nu f_{\nu }\left( u\right) +f_{\nu
-1}\left( u\right) ,  \label{relf}
\end{equation}%
which is obtained by using the recurrence properties of the Macdonald
function \cite{Abra}. For the VEVs of the diagonal components we get (no
summation over $i$)%
\begin{equation}
\langle T_{i}^{i}\rangle =\langle T_{i}^{i}\rangle _{\mathrm{M}}+\sum_{%
\mathbf{n}_{q}}\frac{\omega _{\mathbf{n}_{q}}^{p+2}}{\left( 2\pi \right) ^{%
\frac{p}{2}+1}V_{q}}\int_{0}^{\infty }dy\,\frac{F^{(i)}(y)}{\pi ^{2}/4+y^{2}}%
,  \label{Tii}
\end{equation}%
where $\langle T_{i}^{i}\rangle _{\mathrm{M}}$ is the expectation value in
the Minkowski vacuum and we have defined the functions%
\begin{eqnarray}
F^{(0)}(y) &=&\left( 4\zeta \cosh ^{2}y-1\right) \left[ u^{2}f_{p/2+2}\left(
u\right) -f_{p/2+1}\left( u\right) \right] ,  \notag \\
F^{(1)}(y) &=&\left( 1-4\zeta \cosh ^{2}y\right) f_{p/2+1}\left( u\right) , 
\notag \\
F^{(l)}(y) &=&f_{p/2+1}\left( u\right) +\left( 1-4\zeta \right) \cosh ^{2}y 
\left[ 2f_{p/2+1}\left( u\right) -u^{2}f_{p/2+2}\left( u\right) \right] , 
\notag \\
F^{(r)}(y) &=&F^{(2)}(y)-f_{p/2+1}\left( u\right) +\frac{k_{r}^{2}}{\omega _{%
\mathbf{n}_{q}}^{2}}f_{p/2}\left( u\right) ,  \label{Fr}
\end{eqnarray}%
with $l=2,\ldots ,p+1$, $r=p+2,\ldots ,D$, and $u$ given by (\ref{u}). When
deriving these expressions, the relation 
\begin{equation}
f_{\nu }^{\prime }(u)=-uf_{\nu +1}(u)  \label{frel}
\end{equation}%
has been used. The function $u^{2}f_{p/2+2}\left( u\right) $ in (\ref{Fr})
can be excluded by taking into account the formula (\ref{relf}).

The Minkowskian part in (\ref{Tii}) is given by (see Appendix \ref{sec:Mink})%
\begin{equation}
\langle T_{i}^{i}\rangle _{\mathrm{M}}=-\frac{m^{D+1}}{(2\pi )^{\frac{D+1}{2}%
}}\sideset{}{'}{\sum}_{\mathbf{n}_{q}}\cos (\mathbf{n}_{q}\cdot \boldsymbol{%
\tilde{\alpha}}_{q})G^{(i)}(mg_{\mathbf{n}_{q}}),  \label{TiiM}
\end{equation}%
with the functions%
\begin{eqnarray}
G^{(l)}(u) &=&f_{\frac{D+1}{2}}(u),  \notag \\
G^{(r)}(u) &=&f_{\frac{D+1}{2}}(u)-m^{2}L_{r}^{2}n_{r}^{2}f_{\frac{D+3}{2}%
}(u),  \label{FrM}
\end{eqnarray}%
where $l=0,1,\ldots ,p+1$ and $r=p+2,\ldots ,D$. The energy-momentum tensor (%
\ref{TiiM}) describes a gravitational source with a cosmological
constant-type equation of state in the uncompact subspace, $P_{\mathrm{(M)}%
l}=-\varepsilon _{\mathrm{(M)}}$, with the energy density $\varepsilon _{%
\mathrm{(M)}}=\langle T_{0}^{0}\rangle _{\mathrm{M}}$ and effective pressure 
$P_{\mathrm{(M)}l}=-\langle T_{l}^{l}\rangle _{\mathrm{M}}$ along the
uncompact dimension $x^{l}$. For periodic conditions, $\boldsymbol{\tilde{%
\alpha}}_{q}=\boldsymbol{0}_{q}$, the vacuum energy density is negative and
we have $P_{\mathrm{(M)}r}<P_{\mathrm{(M)}1}$. In particular, in a model
with a single compact dimension $x^{D}$ and for a massless field, one gets $%
P_{\mathrm{(M)}D}=D\varepsilon _{\mathrm{(M)}}$. This type of equation of
state cannot be realized by classical sources.

In addition to the diagonal components, the VEV of the energy-momentum
tensor has nonzero off-diagonal components%
\begin{equation}
\langle T^{il}\rangle =\langle T^{il}\rangle _{\mathrm{M}}-\sum_{\mathbf{n}%
_{q}}\frac{k^{i}k^{l}}{V_{q}}\omega _{\mathbf{n}_{q}}^{p}\int_{0}^{\infty
}dy\,\frac{f_{p/2}\left( 2\xi \omega _{\mathbf{n}_{q}}\cosh y\right) }{%
\left( 2\pi \right) ^{\frac{p}{2}+1}\left( y^{2}+\pi ^{2}/4\right) },
\label{Til}
\end{equation}%
with $i\neq l$ and $i,l=p+2,\ldots ,D$. The Minkowskian VEV in (\ref{Til})
is expressed as%
\begin{equation}
\langle T^{il}\rangle _{\mathrm{M}}=-\frac{m^{D+3}L_{i}L_{l}}{\left( 2\pi
\right) ^{\frac{D+1}{2}}}\sideset{}{'}{\sum}_{\mathbf{n}_{q}}n_{i}n_{l}\cos (%
\mathbf{n}_{q}\cdot \boldsymbol{\tilde{\alpha}}_{q})f_{\frac{D+3}{2}}(mg_{%
\mathbf{n}_{q}}).  \label{TilM2}
\end{equation}%
The off-diagonal components become zero if $\tilde{\alpha}_{i}$ or $\tilde{%
\alpha}_{l}$ are equal to zero. As an additional check, it can be seen that
the relation 
\begin{equation}
\sum_{i=0}^{D}\langle T_{i}^{i}\rangle =\left[ D\left( \zeta -\zeta
_{D}\right) \nabla _{l}\nabla ^{l}+m^{2}\right] \langle \varphi \varphi
^{+}\rangle  \label{Trrel}
\end{equation}%
takes place for the trace of the energy-momentum tensor. In the problem
under consideration, the covariant conservation equation $\nabla
_{l}T_{i}^{l}=0$ is reduced to the single equation $\partial _{\xi }\left(
\xi \langle T_{1}^{1}\rangle \right) =\langle T_{0}^{0}\rangle $. The
validity of this relation is easily verified using this formulas (\ref{relf}%
) and (\ref{frel}). For a conformally coupled massless field the vacuum
energy-momentum tensor is traceless. In the problem under consideration the
background spacetime is flat and the trace anomaly is absent.

Alternative representations for the components of the vacuum energy-momentum
tensor are obtained by using the formula (\ref{SumTrans}). The expressions
for the components $\langle T_{i}^{i}\rangle $ with $i=0,1,\ldots ,p+1$ are
directly obtained from (\ref{Tii}) and (\ref{Fr}) applying to the series
over $\mathbf{n}_{q}$ the formula (\ref{SumTrans}) with $n=1,2$.
Transformations for the components along compact dimensions are described in
Appendix \ref{sec:Alt}. The diagonal components are presented in the form%
\begin{equation}
\langle T_{i}^{i}\rangle =\langle T_{i}^{i}\rangle _{\mathrm{M}}+\frac{%
m^{D+1}}{(2\pi )^{\frac{D+1}{2}}}\sum_{\mathbf{n}_{q}}\cos (\mathbf{n}%
_{q}\cdot \boldsymbol{\tilde{\alpha}})\int_{0}^{\infty }dy\,\frac{%
Z^{(i)}(mw_{\mathbf{n}_{q}})}{y^{2}+\pi ^{2}/4},  \label{Tii2}
\end{equation}%
with the functions%
\begin{eqnarray}
Z^{(0)}(u) &=&\left( 4\zeta \cosh ^{2}y-1\right) \left[ \left( 2m\xi \cosh
y\right) ^{2}f_{\frac{D+3}{2}}(u)-f_{\frac{D+1}{2}}(u)\right] ,  \notag \\
Z^{(1)}(u) &=&\left( 1-4\zeta \cosh ^{2}y\right) f_{\frac{D+1}{2}}(u), 
\notag \\
Z^{(l)}(u) &=&f_{\frac{D+1}{2}}(u)+\left( 1-4\zeta \right) \cosh ^{2}y\left[
2f_{\frac{D+1}{2}}(u)-\left( 2m\xi \cosh y\right) ^{2}f_{\frac{D+3}{2}}(u)%
\right] ,  \notag \\
Z^{(r)}(u) &=&Z^{(2)}(u)-m^{2}L_{r}^{2}n_{r}^{2}f_{\frac{D+3}{2}}(u),
\label{Zr}
\end{eqnarray}%
and $l=2,\ldots ,p+1$, $r=p+2,\ldots ,D$. In (\ref{Tii2}), the replacement (%
\ref{Repl}) can be made. For the off-diagonal components we get%
\begin{equation}
\langle T^{il}\rangle =\langle T^{il}\rangle _{\mathrm{M}}+\frac{%
m^{D+3}L_{i}L_{l}}{(2\pi )^{\frac{D+1}{2}}}\sum_{\mathbf{n}%
_{q}}n_{i}n_{l}\cos (\mathbf{n}_{q}\cdot \boldsymbol{\tilde{\alpha}}%
)\int_{0}^{\infty }dy\,\frac{f_{\frac{D+3}{2}}(mw_{\mathbf{n}_{q}})}{%
y^{2}+\pi ^{2}/4},  \label{Til2b}
\end{equation}%
where $i,l=p+2,\ldots ,D$ and $i\neq l$. In this formula we can make the
replacement 
\begin{equation}
\cos (\mathbf{n}_{q}\cdot \boldsymbol{\tilde{\alpha}})\rightarrow -\sin
(n_{i}\tilde{\alpha}_{i})\sin (n_{l}\tilde{\alpha}_{l})\prod\limits_{j=p+2,%
\neq i,l}^{D}\cos (n_{j}\tilde{\alpha}_{j}).  \label{Reploff}
\end{equation}%
This shows that the VEV $\langle T^{il}\rangle $ is an odd periodic function
of the phases $\tilde{\alpha}_{i}$ and $\tilde{\alpha}_{l}$, and even
periodic functions of the remaining phases, with the period of $2\pi $. This
correspond to the periodicity with respect to the corresponding magnetic
fluxes with the period of flux quantum.

The VEVs in the Rindler spacetime with trivial topology are obtained from (%
\ref{Tii2}) and (\ref{Til2b}) in the limit $L_{i}\rightarrow \infty $. In
this limit the contribution of the terms with $n_{i}=0$, $i=p+2,\ldots ,D$,
only survives and the off-diagonal components vanish. For the corresponding
renormalized VEV in the Minkowski vacuum one has $\langle T_{il}\rangle _{%
\mathrm{M}}=0$, and we get (no summation over $i$) \cite{Saha02} (see also 
\cite{Hill86} for alternative representations in special cases of
conformally and minimally coupled real scalar fields)%
\begin{equation}
\langle T_{i}^{i}\rangle _{0}=\frac{m^{D+1}}{(2\pi )^{\frac{D+1}{2}}}%
\int_{0}^{\infty }dy\,\frac{Z^{(i)}(2m\xi \cosh y)}{y^{2}+\pi ^{2}/4},
\label{TiFR}
\end{equation}%
with the functions $Z^{(i)}(u)$ from (\ref{Zr}) (with $n_{r}=0$ in the
function $Z^{(r)}(u)$). As expected, the stresses in the subspace with the
coordinates $(x^{2},x^{3},\ldots ,x^{D})$ are isotropic, $\langle
T_{i}^{i}\rangle _{0}=\langle T_{2}^{2}\rangle _{0}$ for $i=3,\ldots ,D$. By
taking into account that the part in (\ref{Tii2}) with $\mathbf{n}_{q}=%
\mathbf{0}_{q}$ presents the VEV (\ref{TiFR}), the diagonal components are
decomposed as (no summation over $i$)%
\begin{equation}
\langle T_{i}^{i}\rangle =\langle T_{i}^{i}\rangle _{0}+\langle
T_{i}^{i}\rangle _{\mathrm{t}},  \label{Tidec}
\end{equation}%
where the topological contribution is given by%
\begin{equation}
\langle T_{i}^{i}\rangle _{\mathrm{t}}=\frac{m^{D+1}}{(2\pi )^{\frac{D+1}{2}}%
}\sideset{}{'}{\sum}_{\mathbf{n}_{q}}\cos (\mathbf{n}_{q}\cdot \boldsymbol{%
\tilde{\alpha}})\int_{0}^{\infty }dy\,\frac{Z^{(i)}(mw_{\mathbf{n}%
_{q}})-G^{(i)}(mg_{\mathbf{n}_{q}})}{y^{2}+\pi ^{2}/4}.  \label{Tit}
\end{equation}%
In Rindler spacetime with trivial topology, the off-diagonal component
vanishes and the corresponding non-zero VEV is a purely topological effect.
By taking into account the expression (\ref{TilM2}) for the Minkowski
vacuum, the VEV is presented in the form%
\begin{equation}
\langle T^{il}\rangle =\frac{m^{D+3}L_{i}L_{l}}{(2\pi )^{\frac{D+1}{2}}}%
\sum_{\mathbf{n}_{q}}n_{i}n_{l}\cos (\mathbf{n}_{q}\cdot \boldsymbol{\tilde{%
\alpha}})\int_{0}^{\infty }dy\,\frac{f_{\frac{D+3}{2}}(mw_{\mathbf{n}%
_{q}})-f_{\frac{D+3}{2}}(mg_{\mathbf{n}_{q}})}{y^{2}+\pi ^{2}/4},
\label{Til3}
\end{equation}%
with $i,l=p+2,\ldots ,D$. This shows that the off-diagonal component becomes
zero on the Rindler horizon. Recall that we can make the replacements (\ref%
{Repl}) and (\ref{Reploff}) in (\ref{Tit}) and (\ref{Til}), respectively.
Note that, in accordance with (\ref{frel}), the function $f_{\nu }(u)$ is
monotonically decreasing and the denominator in the integrand of (\ref{Til})
is negative.

For a massless field the expressions for the diagonal components of the
energy-momentum tensor become%
\begin{equation}
\langle T_{i}^{i}\rangle =\langle T_{i}^{i}\rangle _{\mathrm{M}}+\frac{%
\Gamma \left( \frac{D+1}{2}\right) }{2\pi ^{\frac{D+1}{2}}}\sum_{\mathbf{n}%
_{q}}\cos (\mathbf{n}_{q}\cdot \boldsymbol{\tilde{\alpha}})\int_{0}^{\infty
}dy\,\frac{Z_{0}^{(i)}(w_{\mathbf{n}_{q}})}{\left( y^{2}+\pi ^{2}/4\right)
w_{\mathbf{n}_{q}}^{D+1}},  \label{Tii2m0}
\end{equation}%
where the new functions are defiend by%
\begin{eqnarray}
Z_{0}^{(0)}(u) &=&\left( 4\zeta \cosh ^{2}y-1\right) \left[ \left(
D+1\right) \left( 2\xi \cosh y/u\right) ^{2}-1\right] ,  \notag \\
Z_{0}^{(1)}(u) &=&1-4\zeta \cosh ^{2}y,  \notag \\
Z_{0}^{(l)}(u) &=&1+\left( 1-4\zeta \right) \cosh ^{2}y\left[ 2-\left(
D+1\right) \left( 2\xi \cosh y/u\right) ^{2}\right] ,  \notag \\
Z_{0}^{(r)}(u) &=&Z_{0}^{(2)}(u)-\left( D+1\right) L_{r}^{2}n_{r}^{2}/u^{2}.
\label{Zrm0}
\end{eqnarray}%
The off-diagonal component is expressed as%
\begin{equation}
\langle T_{il}\rangle =\langle T_{il}\rangle _{\mathrm{M}}+\frac{L_{i}L_{l}}{%
\pi ^{\frac{D+1}{2}}}\Gamma \left( \frac{D+3}{2}\right) \sum_{\mathbf{n}%
_{q}}n_{i}n_{l}\cos (\mathbf{n}_{q}\cdot \boldsymbol{\tilde{\alpha}}%
)\int_{0}^{\infty }dy\,\frac{w_{\mathbf{n}_{q}}^{-D-3}}{y^{2}+\pi ^{2}/4}.
\label{Til2bm0}
\end{equation}%
For the diagonal components of the Minkowskian VEV we obtain (no summation
over $i$)%
\begin{equation}
\left\langle T_{i}^{i}\right\rangle _{\mathrm{M}}=-\frac{\Gamma \left( \frac{%
D+1}{2}\right) }{2\pi ^{\frac{D+1}{2}}}\sideset{}{'}{\sum}_{\mathbf{n}_{q}}%
\frac{\cos (\mathbf{n}_{q}\cdot \boldsymbol{\tilde{\alpha}})}{g_{\mathbf{n}%
_{q}}^{D+1}}\left( 1-\delta _{(i)}n_{i}^{2}L_{i}^{2}\frac{D+1}{g_{\mathbf{n}%
_{q}}^{2}}\right) ,  \label{TiiMm0}
\end{equation}%
where $\delta _{(i)}=0$ for $i=0,1,\ldots ,p+1$, and $\delta _{(i)}=1$ for $%
i=p+2,\ldots ,D$. The off-diagonal component becomes%
\begin{equation}
\langle T^{il}\rangle _{\mathrm{M}}=-\frac{L_{i}L_{l}}{\pi ^{\frac{D+1}{2}}}%
\Gamma (\frac{D+3}{2})\sideset{}{'}{\sum}_{\mathbf{n}_{q}}\frac{n_{i}n_{l}}{%
g_{\mathbf{n}_{q}}^{D+3}}\cos (\mathbf{n}_{q}\cdot \boldsymbol{\tilde{\alpha}%
}_{q}),  \label{TilMm0}
\end{equation}%
with $i\neq l$, $i,l=p+2,\ldots ,D$. The vacuum energy-momentum tensor in
the Rindler spacetime with trivial topology is diagonal and the expressions
for the components in the case of a massless field are obtained from (\ref%
{TiFR}) in the limit $m\rightarrow 0$ (no summation over $i$) \cite{Saha02}:%
\begin{equation}
\langle T_{i}^{i}\rangle _{0}=\frac{\Gamma (\frac{D+1}{2})}{2\left( 4\pi
\right) ^{\frac{D+1}{2}}\xi ^{D+1}}\int_{0}^{\infty }dy\,\frac{Z_{\mathrm{R}%
}^{(i)}(y)}{y^{2}+\pi ^{2}/4}\frac{1}{\cosh ^{D+1}y},  \label{Tii00}
\end{equation}%
where the functions in the integrand are defined by%
\begin{eqnarray}
Z_{\mathrm{R}}^{(1)}(y) &=&-\frac{1}{D}Z_{\mathrm{R}}^{(0)}(y)=1-4\zeta
\cosh ^{2}y,  \notag \\
Z_{\mathrm{R}}^{(l)}(y) &=&1+\left( D-1\right) \left( 4\zeta -1\right) \cosh
^{2}y,  \label{ZlR}
\end{eqnarray}%
with $l=2,\ldots ,D$.

In order to find the asymptotic of the diagonal components of the vacuum
energy-momentum tensor near the Rindler horizon, we use the representation (%
\ref{Tii2}). The leading contribution comes from the term $\mathbf{n}_{q}=%
\mathbf{0}_{q}$ which coincides with (\ref{TiFR}). Assuming $m\xi \ll 1$,
the leading term is given by the right-hand side of (\ref{Tii00}) and the
VEVs behave like $1/\xi ^{D+1}$. In the remaining part of the series over $%
\mathbf{n}_{q}$ in (\ref{Tii2}), in the leading order, we can put $\xi =0$.
The mode with $\mathbf{n}_{q}=\mathbf{0}_{q}$ does not contribute to the VEV
of the off-diagonal components and the leading term is found by directly
putting $\xi =0$ in (\ref{Til2b}). With this substitution the function $f_{%
\frac{D+3}{2}}(mw_{\mathbf{n}_{q}})$ becomes $f_{\frac{D+3}{2}}(mg_{\mathbf{n%
}_{q}})$. Evaluating the integral over $y$, we can see that the second term
in right-hand side of(\ref{Til2b}) gives $-\langle T^{il}\rangle _{\mathrm{M}%
}$ which cancels the first term. From here we conclude that the off-diagonal
component of the vacuum energy-momentum tensor tends to zero on the Rindler
horizon.

For large values of $\xi $, corresponding to small accelerations, the
asymptotic of the difference in the VEVs for the Minkowski and
Fulling-Rindler vacua is found analogously to our treatment of the field
squared. The leading term is given by (no summation over $i$) 
\begin{equation}
\langle T_{i}^{i}\rangle \approx \langle T_{i}^{i}\rangle _{\mathrm{M}}+%
\frac{\left( 4\zeta -1\right) \left( \omega _{0}/\xi \right) ^{\frac{p}{2}%
+1}e^{-2\omega _{0}\xi }}{2^{p+1}\pi ^{\frac{p}{2}+2}V_{q}\left( -2\omega
_{0}\xi \right) ^{\delta _{1i}}},  \label{Tiilarge}
\end{equation}%
where $\omega _{0}$ is given by (\ref{om0n}). For the off-diagonal
components we get%
\begin{equation*}
\langle T^{il}\rangle \approx \langle T^{il}\rangle _{\mathrm{M}}-\frac{1}{%
2^{p+1}\pi ^{\frac{p}{2}+2}}\sum_{\mathbf{n}_{q}}\frac{k^{i}k^{l}}{V_{q}}%
\frac{\omega _{\mathbf{n}_{q}}^{p}e^{-2\xi \omega _{\mathbf{n}_{q}}}}{\left(
\xi \omega _{\mathbf{n}_{q}}\right) ^{\frac{p}{2}+1}}.
\end{equation*}%
The dominant contribution to the sum over $\mathbf{n}_{q}$ comes from the
terms with $n_{r}=0$, $r\neq i,l$, and $i,l=\pm 1$. The suppression for the
off-diagonal components is stronger compared with the diagonal components.

\begin{figure}[tbph]
\begin{center}
\begin{tabular}{cc}
\epsfig{figure=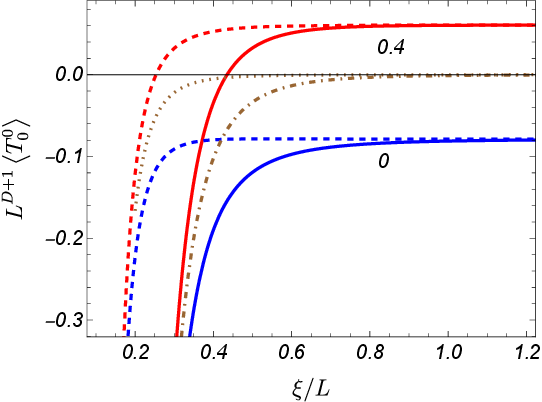,width=8cm,height=6.5cm} & \quad %
\epsfig{figure=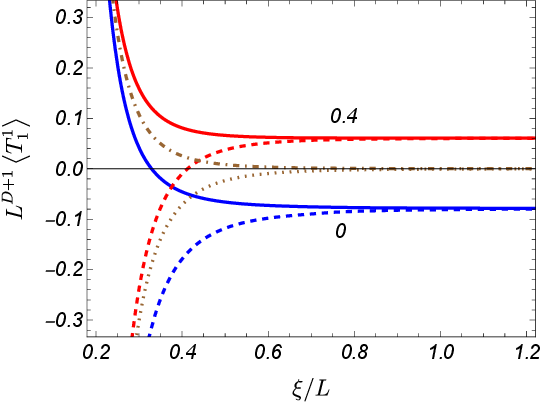,width=8cm,height=6.5cm}%
\end{tabular}%
\end{center}
\caption{The vacuum energy density (left panel) and $_{1}^{1}$-stress for a
massles scalar field versus the ratio $\protect\xi /L$. The full and dashed
curves correspond to minimally and conformally coupled fields and the
numbers near the curves are the values of $\tilde{\protect\alpha}_{D}/2%
\protect\pi $. The dot-dashed and dotted curves present the corresponding
VEVs in the Rindler spacetime without compact dimensions.}
\label{fig3}
\end{figure}

\begin{figure}[tbph]
\begin{center}
\begin{tabular}{cc}
\epsfig{figure=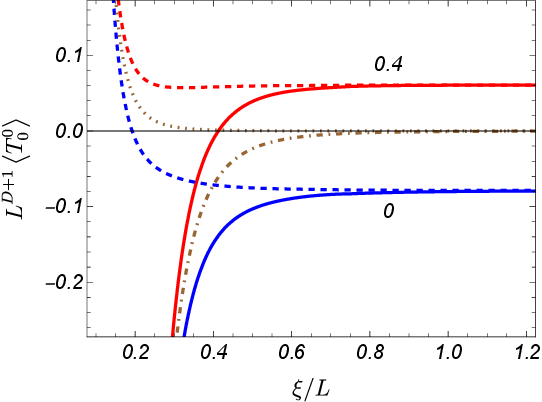,width=8cm,height=6.5cm} & \quad %
\epsfig{figure=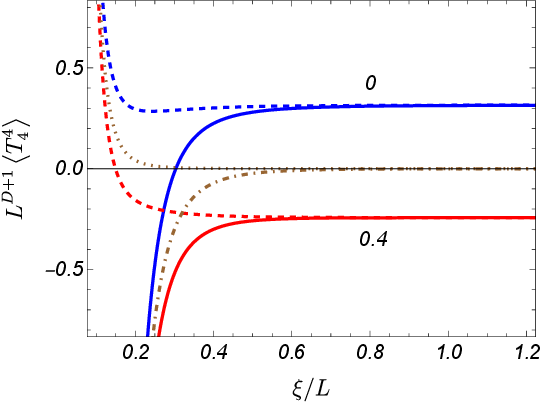,width=8cm,height=6.5cm}%
\end{tabular}%
\end{center}
\caption{The same as in Fig. \protect\ref{fig3} for the stresses $\langle
T_{2}^{2}\rangle $ (left panel) and $\langle T_{4}^{4}\rangle $ (right
panel).}
\label{fig4}
\end{figure}

In figures below the graphs are plotted for the model $D=4$ with a single
compact dimension of the length $L$. First we consider the VEVs for a
massless scalar field. In this special case the combination $L^{D+1}\langle
T_{i}^{i}\rangle $ depends on $\xi $ and $L$ through the ratio $\xi /L$.
Figure \ref{fig3} presents the corresponding dependence for the vacuum
energy density (left panel) and for the stress $\langle T_{1}^{1}\rangle $
(right panel). The numbers near the curves correspond to the values of $%
\tilde{\alpha}_{D}/2\pi $ and the full and dashed curves present the VEVs
for minimally and conformally coupled fields, respectively. The dot-dashed
and dotted curves present the VEVs in $(4+1)$-dimensional Rindler spacetime
without compact dimensions for minimal and conformal couplings,
respectively. Figure \ref{fig4} displays similar dependences for the
stresses $\langle T_{2}^{2}\rangle =\langle T_{3}^{3}\rangle $ and $\langle
T_{4}^{4}\rangle $.

In accordance with the asymptotic analysis given above, for large values of $%
\xi /L$, corresponding to small accelerations or small values of the length
of the compact dimension, the vacuum energy-momentum tensor is approximated
by the corresponding VEV in the Minkowski vacuum. Near the horizon or for
large length of the compact dimension, the the ratio $\xi /L$ is small and
the VEVs $\langle T_{i}^{i}\rangle $ tends to the corresponding results in
Rindler spacetime with trivial topology.

\begin{figure}[tbph]
\begin{center}
\begin{tabular}{cc}
\epsfig{figure=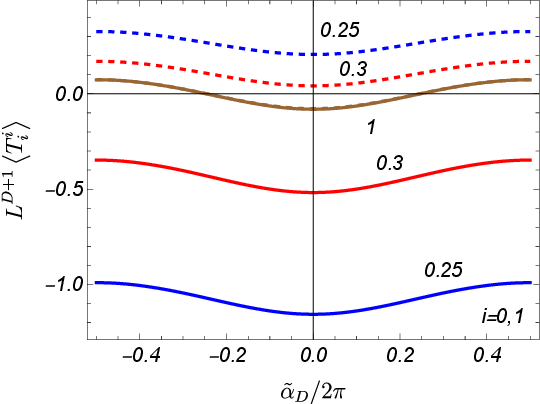,width=8cm,height=6.5cm} & \quad %
\epsfig{figure=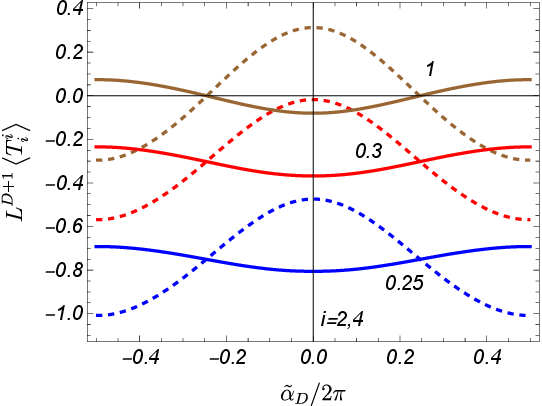,width=8cm,height=6.5cm}%
\end{tabular}%
\end{center}
\caption{The vacuum energy-momentum tensor $\langle T_{i}^{i}\rangle $ as a
functions of $\tilde{\protect\alpha}_{D}/2\protect\pi $ for a minimally
coupled massless field in $4+1$-dimensional Rindler spacetime with a single
compact dimension. The left and right panels correspond to the components $%
i=0,1$ and $i=2,4$, and the numbers near the curves are the values of the
ratio $\protect\xi /L$. The full and dashed curves correspond to $i=0,2$ and 
$i=1,4$, respectively.}
\label{fig5}
\end{figure}

The vacuum energy density and the stresses versus the parameter $\tilde{%
\alpha}_{D}/2\pi $ are plotted in Fig. \ref{fig5} for a minimally coupled
massless field in spatial dimension $D=4$. The VEVs are periodic functions
of this ratio with the period 1 and the graphs are plotted in the interval $|%
\tilde{\alpha}_{D}/2\pi |\leq 0.5$. The left and right panels in Fig. \ref%
{fig5} present the graphs for the components $\langle T_{i}^{i}\rangle $
with $i=0,1$ and $i=2,4$, respectively. The numbers near the curves
correspond to the values of $\xi /L$. Full and dashed curves correspond to
the components $i=0,2$ and $i=1,4$, respectively. For large values of $\xi
/L $ the parts induced by nonzero acceleration are small and the VEVs tend
to the corresponding expectation values for the Minkowski vacuum.

\begin{figure}[tbph]
\begin{center}
\begin{tabular}{cc}
\epsfig{figure=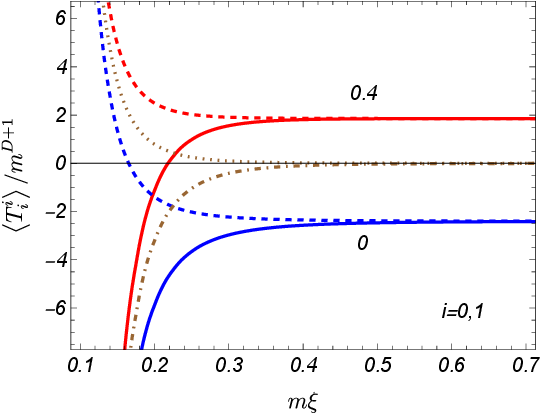,width=8.cm,height=6.5cm} & \quad %
\epsfig{figure=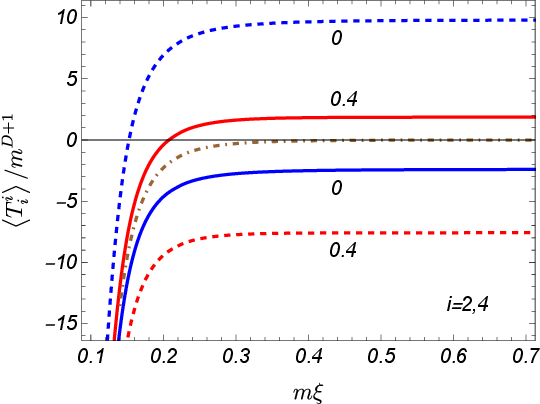,width=8.cm,height=6.5cm}%
\end{tabular}%
\end{center}
\caption{The components of the vacuum energy-momentum tensor for a massive
field versus $m\protect\xi $. The left and right panels correspond to the
components $\langle T_{i}^{i}\rangle $ with $i=0,1$ and $i=2,4$,
respectively. The full and dashed curves present the VEVs for $i=0,2$ and $%
i=1,4$, respectively. The graphs are plotted for $\tilde{\protect\alpha}%
_{D}/2\protect\pi =0,0.4$ (the numbers near the curves) and for $mL=0.5$.
The dot-dashed and dotted curves present the VEVs in the Rindler spacetime
without compact dimensions. }
\label{fig6}
\end{figure}

The VEVs $\langle T_{i}^{i}\rangle $ for a massive scalar field in spatial
dimension $D=4$ are plotted in Fig. \ref{fig6} as functions of $m\xi $. The
left and right panels correspond to the components $i=0,1$ and $i=2,4$,
respectively. The full curves present the components $i=0,2$ and the dashed
curve correspond to $i=1,4$. The numbers near the curves are the values of $%
\tilde{\alpha}_{D}/2\pi $ and the graphs are plotted for $mL=0.5$. The
dot-dashed and dotted curves on the left panel correspond to the VEVs $%
\langle T_{i}^{i}\rangle _{0}$ with $i=0$ and $i=1$, respectively.
Similarly, the dot-dashed curve on the right panel presents the VEVs $%
\langle T_{2}^{2}\rangle _{0}=\langle T_{4}^{4}\rangle _{0}$.

\begin{figure}[tbph]
\begin{center}
\begin{tabular}{cc}
\epsfig{figure=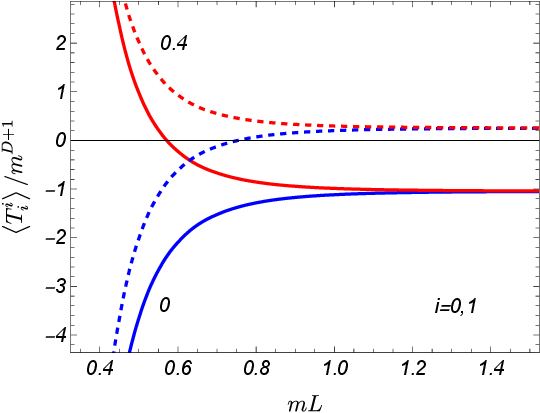,width=8.cm,height=6.5cm} & \quad %
\epsfig{figure=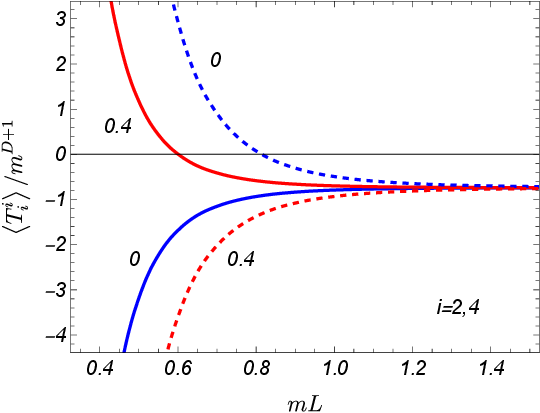,width=8.cm,height=6.5cm}%
\end{tabular}%
\end{center}
\caption{The components of the vacuum energy-momentum tensor as functions of 
$mL$. The left and right panels present the graphs for the components $i=0,1$
and $i=2,4$, respectively. The full and dashed curves correspond to $i=0,2$
and $i=1,4$, respectively. For the other parameters we have taken $\tilde{%
\protect\alpha}_{D}/2\protect\pi =0,0.4$ (the numbers near the curves) and $m%
\protect\xi =0.5$. }
\label{fig7}
\end{figure}

In Fig. \ref{fig7}, for a minimally coupled massive field, we have plotted
the dependence of the VEVs $\langle T_{i}^{i}\rangle $ on the length of the
compact dimension (in units of the Compton wavelength $1/m$). The left and
right panel present the graphs for $i=0,1$ and $i=2,4$, respectively. The
full curves correspond to $i=0,2$ and and for the dashed curves one has $%
i=1,4$. The graphs are plotted for $\tilde{\alpha}_{D}/2\pi =0,0.4$ (the
numbers near the curves) and for $m\xi =0.25$. For large values of the
length of the compact dimension, the topological contributions are small and
the VEVs are dominated by the parts corresponding to Rindler spacetime with
trivial topology. In the opposite limit of small values of $L$, the VEVs are
approximated by the corresponding quantities in the Minkowski vacuum.

\section{Vacuum densities in curved backgrounds with compact dimensions}

\label{sec:Cyl}

In this section, the results obtained above are applied to study the vacuum
densities in curved backgrounds with compact dimensions. We will begin the
discussion with the expectation values near the horizon of non-rotating and
uncharged cylindrical black holes and topological black holes with toroidal
horizons. Some approximate and numerical results for the VEVs of the field
squared and energy-momentum tensor for a massless real scalar field in the
geometry of 4-dimensional cylindrical black hole are presented in \cite%
{Bene98,Bene99}. The cylindrical black holes are axially symmetric solutions
of the Einstein equations with negative cosmological constant $\Lambda $.
The exterior geometry is covered by the coordinates $x_{\mathrm{bh}%
}^{i}=(t,r,\mathbf{y},\mathbf{\phi })$ with dimensionless sets $\mathbf{y}%
=(y^{1},\ldots ,y^{p})$, $\mathbf{\phi }=(\phi _{1},\ldots ,\phi _{q})$, and 
$p+q+1=D$. For the coordinates $y^{i}$ we have $-\infty <y^{i}<+\infty $ and
the angular coordinates $\phi _{i}$ in the compact subspace vary in the
range $0\leq \phi _{i}<2\pi $. The corresponding metric tensor is given by
the line element (for generalizations in the cases of rotating and charged
black holes see \cite{Awad03}) 
\begin{equation}
ds_{\mathrm{bh}}^{2}=f(r)dt^{2}-\frac{dr^{2}}{f(r)}-r^{2}\left( d\mathbf{y}%
\right) ^{2}-r^{2}\left( d\mathbf{\phi }\right) ^{2},  \label{ds2bh}
\end{equation}%
where the radial function is defined by 
\begin{equation}
f(r)=\frac{r^{2}}{a^{2}}-\frac{M}{r^{D-2}},\;a=\sqrt{D\frac{D-1}{-2\Lambda }}%
.  \label{fr}
\end{equation}%
For the parameter $M$ one has $M=16\pi G_{D+1}\mathcal{M}/(D-1)$, where $%
G_{D+1}$ is the Newton gravitational constant and $\mathcal{M}$ is the mass
per unit volume of the subspace covered by the coordinates $\mathbf{y}$. The
cylindrical black holes have been considered in different contexts,
including cosmic strings, supergravity and low energy string theories.

The event horizon for the line element (\ref{ds2bh}) is located at $%
r=r_{H}\equiv \left( a^{2}M\right) ^{1/D}$. Near the horizon, we pass to the
radial coordinate $\xi $ defined according to 
\begin{equation}
\xi =\frac{2a}{\sqrt{D}}\sqrt{r/r_{H}-1}.  \label{ksibh}
\end{equation}
Introducing the coordinates $x^{i}$ with $x^{0}=\tau $, $x^{1}=\xi $, and 
\begin{eqnarray}
\tau &=&\frac{D\sqrt{M}}{2ar_{H}^{D/2-1}}t,\;x^{i}=r_{H}y^{i-1},\;i=2,\ldots
,p+1  \notag \\
x^{l} &=&r_{H}\phi _{l-p-1},\;l=p+2,\ldots ,D,  \label{Horxi}
\end{eqnarray}%
the line element $ds_{\mathrm{bh}}^{2}$ is approximated by the Rindler line
element (\ref{ds2}). The compact subspace with toroidal topology is covered
by the coordinates $(x^{p+2},\ldots ,x^{D})$ with the lengths $L_{l}=2\pi
r_{H}$. The VEVs of the field squared and energy-momentum tensor in the
near-horizon region of cylindrical black holes are expressed by the formulas
given above with $\xi $ given by (\ref{ksibh}). The approximation works
under the condition $\left( \xi /a\right) ^{2}\ll 1$, where the length scale 
$a$ is determined by the negative cosmological constant. The second length
scale $r_{G}$ is determined by the mass parameter according to $%
r_{G}=M^{1/(D-2)}$. The condition of the near-horizon approximation for the
ratio $\xi /L_{l}$ is presented as $\xi /L_{l}\ll \left( a/r_{G}\right)
^{1-2/D}/(2\pi )$. For $a$ and $r_{G}$ having the same order of magnitude
one has $\xi /L_{l}\ll 1$. Consequently, the simple asymptotic formulas for
the VEVs can be utilized. For black holes with $a\gg r_{G}$ we can have $\xi
/L_{l}\gtrsim 1$. In such cases, it is necessary to employ exact expressions.

Another approximation can be used in the region $r\gg r_{H}$ corresponding
to large distances from the horizon. With new coordinates 
\begin{eqnarray}
z &=&a^{2}/r,\;x^{i}=ay^{i-1},\;i=2,\ldots ,p+1,  \notag \\
x^{l} &=&a\phi _{l-p-1},\;l=p+2,\ldots ,D,  \label{LargeDist}
\end{eqnarray}%
the line element (\ref{ds2bh}) is approximated by%
\begin{equation}
ds_{\mathrm{bh}}^{2}\approx \frac{a^{2}}{z^{2}}\left[ dt^{2}-dz^{2}-%
\sum_{i=2}^{D}\left( dx^{i}\right) ^{2}\right] ,  \label{ds2bh2}
\end{equation}%
where the variation ranges of the coordinates $x^{i}$ are the same as in the
problem we have considered for the Rindler spacetime with lengths of compact
dimensions $L_{l}=2\pi a$, $l=p+2,\ldots ,D$. The line element corresponds
to locally AdS spacetime in Poincar\'{e} coordinates with a compact
subspace. In terms of the coordinate $z$, the large distance condition $r\gg
r_{H}$ is translated as $z/a\ll (a/r_{G})^{1-2/D}$. If the length scales $a$
and $r_{G}$ are of the same order of magnitude, this condition corresponds
to the region near the AdS boundary located at $z=0$. The effects of the
nontrivial topology for the VEVs of the field squared and energy-momentum
tensor on the background geometry described by the right-hand side of (\ref%
{ds2bh2}) can be studied by using the Hadamard function considered in \cite%
{Beze15}.

It should be noted that various vacuum states for black holes are discussed
in the literature. The expectation values obtained from the VEVs derived in
the previous sections correspond to a vacuum state that is locally
equivalent to the Fulling-Rindler vacuum and is known as the Boulware
vacuum. The Boulware vacuum is empty with respect to the static Killing time 
$t$ in (\ref{ds2bh}). It is the natural vacuum for static observers far from
the gravitating object. While the Boulware vacuum is a sensible choice for a
static configuration with no horizon, for a black hole its energy-momentum
tensor diverges at the horizon. A more appropriate for a black hole formed
by collapse is the Unruh vacuum. The corresponding expectation values are
regular on the future horizon. In the Unruh state, static observers at
infinity detect Hawking flux of thermal radiation. For the expectation
values of the energy-momentum tensors in the Unruh and Boulware vacua the
relation 
\begin{equation}
\langle T_{il}\rangle _{\mathrm{U}}=\langle T_{il}\rangle _{\mathrm{B}%
}+\langle T_{il}\rangle _{\mathrm{th}}  \label{RelTUB}
\end{equation}%
holds, where the second term in the right-hand side is the contribution of
the thermal radiation. At points outside the horizon, the expectation values 
$\langle T_{il}\rangle _{\mathrm{U}}$ and $\langle T_{il}\rangle _{\mathrm{B}%
}$ are renormalized by the same prescription and the state-dependent
difference $\langle T_{il}\rangle _{\mathrm{th}}$ is finite at these points.
It describes the change in the stress-energy tensor resulting from a change
of quantum state. At the horizon, the Boulware divergence is canceled by the
state-dependent contribution. A relation similar to (\ref{RelTUB}) holds for
the Hartle-Hawking and Boulware vacua as well.

The above transformation regarding the Rindler spacetime approximation can
be generalized to a static cylindrical object of radius $r_{0}>r_{H}$, whose
geometry in the region $r>r_{0}$ is described by the metric (\ref{ds2bh}).
Near the boundary of the configuration, assuming that $r/r_{0}-1\ll 1$, we
introduce new coordinates%
\begin{equation}
\tau =\frac{f_{1}t}{2r_{0}},\;\xi =\frac{2r_{0}}{\sqrt{f_{1}}}\sqrt{\frac{r}{%
r_{0}}-1+\frac{f_{0}}{f_{1}}},  \label{tauxi}
\end{equation}%
and the coordinates $x^{i}$, $i=2,3,\ldots ,D$, according to (\ref{Horxi})
with the replacement $r_{H}\rightarrow r_{0}$, where%
\begin{equation}
f_{0}=f(r_{0}),\;f_{1}=2f_{0}+\frac{DM}{r_{0}^{D-2}}.  \label{f01}
\end{equation}%
The line element (\ref{ds2bh}) is approximated by (\ref{ds2}). In the static
configuration under consideration, the Killing vector $\partial _{t}$ is
timelike everywhere and the particle creation does not take place. The
Boulware vacuum is the most natural vacuum state, being the lowest energy
stable state (see \cite{Frol98} for a discussion in the Schwarzschild
geometry).

Using the results obtained above, we can generate vacuum expectation values
for a conformally coupled massless field ($\zeta =\zeta _{D}$) in
conformally flat spacetimes whose metric is expressed in terms of local
Rindler coordinates. For a conformally flat spacetime with the line element $%
ds_{\mathrm{CF}}^{2}=\Omega _{\mathrm{M}}^{2}(x_{\mathrm{M}}^{k})\eta
_{il}dx_{\mathrm{M}}^{i}dx_{\mathrm{M}}^{l}$, switching to the Rindler
coordinates according to (\ref{txM}), the interval becomes 
\begin{equation}
ds_{\mathrm{CF}}^{2}=\Omega ^{2}(x^{k})ds^{2},  \label{ds2CF}
\end{equation}%
where $ds^{2}$ is given by (\ref{ds2}) and the conformal factor $\Omega
^{2}(x^{k})$ is obtained from $\Omega _{\mathrm{M}}^{2}(x_{\mathrm{M}}^{k})$
with the coordinate substitutions from (\ref{txM}). Important examples
include Friedmann-Robertson-Walker cosmological models with flat spatial
sections and compact spatial dimensions ($\Omega _{\mathrm{M}}^{2}(x_{%
\mathrm{M}}^{k})=\Omega _{\mathrm{M}}^{2}(t_{\mathrm{M}})$) , as well as the
patch of the anti-de Sitter (AdS) spacetime covered by Poincar\'{e}
coordinates ($\Omega _{\mathrm{M}}^{2}(x_{\mathrm{M}}^{k})=\alpha
^{2}/(x^{2})^{2}$, $0\leq x^{2}<\infty $) . Let us denote by $\langle \cdots
\rangle _{\mathrm{CFR}}$ the expectation values in the geometry (\ref{ds2CF}%
) for the state conformally related to the Fulling-Rindler vacuum in flat
spacetime. Then, from the standard relation between the expectation values
in conformally related problems we get 
\begin{equation}
\langle \bar{T}_{i}^{l}\rangle _{\mathrm{CFR}}=\frac{\langle
T_{i}^{l}\rangle }{\Omega ^{D+1}}+\langle T_{i}^{l}\rangle _{\mathrm{g}},
\label{TCR}
\end{equation}%
where the expectation value $\langle T_{i}^{l}\rangle $ is given by the
formulas in the previous section, while the part $\langle T_{i}^{l}\rangle _{%
\mathrm{g}}$ is purely geometrical. The latter contribution is determined by
the local geometrical characteristics of the background spacetime.

As with the geometry considered in the previous section, the spacetime
described by the line element (\ref{ds2CF}) has a spatial topology $%
R^{p}\times (S^{1})^{q}$. The important point here is that the local
geometric characteristics are the same for the topologies $R^{p}\times
(S^{1})^{q}$ and $R^{p+q}$, and hence the geometrical contribution $\langle 
\bar{T}_{\mu }^{\nu }\rangle _{\mathrm{g}}$ is also the same for these
topologies. This means that the topological contribution, which depends on
the lengths of the compact dimensions and the phases $\tilde{\alpha}_{l}$,
is contained in the first term on the right-hand side of (\ref{TCR}). Thus,
the above formulas allow us to study the effects of nontrivial topology in
the conformally flat geometries for the state that is conformally related to
the Fulling-Rindler vacuum state. Among the geometries (\ref{ds2CF}), the
Rindler-AdS spacetime has attracted a special attention as a controlled
theoretical framework for investigating the thermal properties of quantum
states, the holographic entanglement entropy, and the relationship between
the bulk geometry and the boundary quantum field theory in the AdS/CFT
correspondence (see \cite{Empa99}-\cite{Zhan23} and the references therein).
Note that, for a $(D+1)$-dimensional AdS bulk (\ref{ds2CF}), the background
geometry of the dual conformal field theory living on the AdS boundary is
described by the $D$-dimensional Rindler line element obtained from (\ref%
{ds2}) by excluding the contribution coming from the coordinate normal to
the AdS boundary ($x^{2}$ coordinate). In the presence of compact
dimensions, the expressions for the corresponding VEVs can be obtained from
the above formulas by replacing $p$ with $p-1$.

An interesting example of a (2+1)-dimensional curved spacetime that is
conformally related to a Rindler spacetime with a single compact dimension $%
x^{2}$ ($p=q=1$), is the Beltrami pseudosphere, which is described by the
line element 
\begin{equation*}
ds_{\mathrm{BP}}^{2}=\frac{a^{2}}{\xi ^{2}}\left( \xi ^{2}d\tau ^{2}-d\xi
^{2}-L^{2}d\phi ^{2}\right) ,
\end{equation*}%
where $0\leq \phi \leq 2\pi $ and $a$ is the curvature radius. The conformal
relation with the (2+1)-dimensional Rindler line element (\ref{ds2}) becomes 
$ds_{\mathrm{BP}}^{2}=\left( a/\xi \right) ^{2}ds^{2}$, where for the
compact dimension $x^{2}$ one has $x^{2}=L\phi $. From here it follows that $%
L_{2}=2\pi L$. In this special case and for a conformally coupled massless
scalar field, the topological contribution in the VEV of the field squared
is obtained from (\ref{phi2m0}) with $D=2$, excluding the term with $\mathbf{%
n}_{q}=n_{2}=0$. It is expressed as%
\begin{equation}
\langle \varphi \varphi ^{\dagger }\rangle _{\mathrm{t}}=\sum_{n=1}^{\infty }%
\frac{\cos (n\tilde{\alpha}_{2})}{2\pi \xi }\left[ \frac{1}{u_{n}}%
-\int_{0}^{\infty }\,dy\frac{\left( y^{2}+\frac{\pi ^{2}}{4}\right) ^{-1}}{%
\sqrt{4\cosh ^{2}y+u_{n}^{2}}}\right] ,  \label{phi2tD2}
\end{equation}%
where $u_{n}=nL_{2}/\xi $.

The topological contributions in the VEVs of the field squared and the
energy-momentum tensor for a massive scalar field with a general curvature
coupling parameter localized on the Beltrami pseudosphere are studied in 
\cite{Petr26} (see \cite{Saha24} for the VEV\ of the current density). For a
conformally coupled massless field the corresponding formula is reduced to 
\begin{equation}
\langle \varphi \varphi ^{\dagger }\rangle _{\mathrm{BP,t}%
}=\sum_{n=1}^{\infty }\frac{\cos (n\tilde{\alpha}_{2})}{\pi ^{2}a}%
\int_{0}^{\infty }d\nu \,Q_{\nu -\frac{1}{2}}(1+u_{n}^{2}/2),  \label{phi2BP}
\end{equation}%
where $Q_{\mu }(x)$ is the Legendre function of the second kind. The
conformal relation $\langle \varphi \varphi ^{\dagger }\rangle _{\mathrm{BP,t%
}}=(\xi /a)\langle \varphi \varphi ^{\dagger }\rangle _{\mathrm{t}}$ is
expected between the VEVs (\ref{phi2tD2}) and (\ref{phi2BP}). For $u>0$, we
have numerically checked the equality 
\begin{equation}
\frac{2}{\pi }\int_{0}^{\infty }d\nu \,Q_{\nu -\frac{1}{2}}\left( 1+\frac{1}{%
2}u^{2}\right) =\frac{1}{u}-\int_{0}^{\infty }\,dy\frac{\left( y^{2}+\frac{%
\pi ^{2}}{4}\right) ^{-1}}{\sqrt{4\cosh ^{2}y+u^{2}}},  \label{Rel3}
\end{equation}%
which shows that the relationship does indeed take place. The conformal
relation $\langle T_{i}^{l}\rangle _{\mathrm{BP,t}}=(\xi /a)^{3}\langle
T_{i}^{l}\rangle _{\mathrm{t}}$ can also be verified for the topological
contributions to the VEVs of the energy-momentum tensors. Realizations of
the Beltrami pseudosphere and (2+1)-dimensional Rindler spacetime as
effective geometries for quasiparticles in 2D materials such as graphene
have been discussed in \cite{Iori12,Iori16,Bhar23}. An effective geometry
can be engineered using strain, a temperature gradient, or by depositing the
material over a substrate with a given shape.

\section{Conclusion}

\label{sec:Conc}

We studied the impact of compactifying a part of spatial dimensions in
Rindler spacetime on the local properties of the FR vacuum state for a
charged massive scalar field with a general curvature coupling parameter.
General quasi-periodicity conditions (\ref{PC}) are imposed along the
compact dimensions. The phases of these conditions are interpreted in terms
of the magnetic flux enclosed by the compact dimensions, and vice versa. The
expectation values of the field squared and the energy-momentum tensor are
considered as important characteristics. To evaluate these VEVs, we use the
Hadamard function representation (\ref{Had}). This representation explicitly
shows the difference in the Hadamard functions for the FR and Minkowski
vacua. This representation is important because the renormalization of the
Rindler VEVs reduces to the renormalisation of the Minkowskian VEVs.

The VEV of the field squared is decomposed as (\ref{phi2}), where the
corresponding renormalized VEV in the locally Minkowski spacetime with
compact dimensions is given by (\ref{phi2M}). The difference between the
VEVs corresponding to the FR and Minkowski vacua is negative. An alternative
representation of the VEV $\langle \varphi \varphi ^{\dagger }\rangle $ is
provided by the formula (\ref{phi22}). The term with $\mathbf{n}_{q}=\mathbf{%
0}_{q}$ in this representation corresponds to the VEV in the Rindler
spacetime with trivial topology (see (\ref{phi2FR})). As a result, the
representation (\ref{phi22}) allows to separate explicitly the part in the
VEV induced by nontrivial topology, expressed by (\ref{phi2t}). For large
accelerations, corresponding to small values of the coordinate $\xi $, the
leading term in the asymptotic expansion over $1/\xi $ coincides with the
VEV for a massless scalar field in the Rindler spacetime without compact
dimensions. This shows that near the Rindler horizon the effects of
nontrivial topology are weak. For small accelerations (large values of $\xi $%
), the leading term in the corresponding expansion coincides with the VEV in
the Minkowski vacuum. The correction induced by acceleration is suppressed
by the factor $e^{-2\xi \omega _{0}}$ with $\omega _{0}$ defined by (\ref%
{om0n}). The exception is the case of a massless field with zero values of
the phases $\tilde{\alpha}_{i}$. In this special case, the decay of the
acceleration-induced contribution follows a power law of the form $1/\xi
^{p} $ (see (\ref{phi2largem0})). When the lengths of the compact dimensions
are small compared to both $\xi $ and $1/m$, the Minkowskian part dominates
in the VEV and the leading contribution is described by (\ref{phi2m0M}).

Another important characteristic of the vacuum is the expectation value of
the energy-momentum tensor. For the diagonal components, its decomposition
into Minkowskian vacuum and acceleration-induced contributions is given by (%
\ref{Tii}) with the functions $F^{(i)}(y)$ defined by (\ref{Fr}). In
addition to the diagonal components, the vacuum energy-momentum tensor also
has off-diagonal components with the indices in the compact subspace and
given by the formula (\ref{Til}). In these representations the Minkowskian
parts of the VEVs are separated. For general values of the phases along the
compact dimensions, these parts are considered in Appendix \ref{sec:Mink}.
Another representation of the vacuum energy-momentum tensor components is
given in (\ref{Tii2}) and (\ref{Til2b}). The VEVs in the Rindler spacetime
without compact dimensions, previously considered in the literature, are
obtained in the decopmactification limit. The effects of the nontrivial
topology are included in the topological contribution, as expressed by (\ref%
{Tit}). To provide an additional check of the results obtained, it was
demonstrated that the vacuum energy-momentum tensor satisfies the covariant
conservation equation and the trace equation (\ref{Trrel}). In particular,
the trace of the energy-momentum tensor becomes zero for a conformally
coupled massless field. Near the horizon and for the diagonal components,
the leading terms in the expansion over $\xi $ coincide with the
corresponding VEVs in the Rindler spacetime without compact dimensions for a
massless field (see (\ref{Tii00})) and these components behave like $1/\xi
^{D+1}$. The off-diagonal components vanish on the Rindler horizon. For
large values of $\xi $, the contribution induced by acceleration decays
exponentially, like $e^{-2\xi \omega _{0}}/\xi ^{p/2+1}$ (with an additional
factor of $1/\xi $ for the component $\langle T_{1}^{1}\rangle $) for
diagonal components. The fall-off of the off-diagonal components is
stronger. In the special case of $\omega _{0}=0$, the off-diagonal
components become zero and one has power-law decay for the acceleration
induced parts in the diagonal components.

The obtained VEVs are applied to study vacuum densities in curved spacetimes
with nontrivial topology. Examples include cylindrical configurations with
toroidal dimensions and curved backgrounds conformally related to the
Rindler spacetime with nontrivial topology. The latter include cosmological
spacetimes and locally AdS spacetimes with toroidal dimensions and curved
tubes of two-dimensional materials.

\section*{Acknowledgments}

The work was supported by the grant No. 21AG-1C047 of the Higher Education
and Science Committee of the Ministry of Education, Science, Culture and
Sport RA. G. V. Mirzoyan was supported by the grant No. 21AG-1C006 of the
Higher Education and Science Committee of the Ministry of Education,
Science, Culture and Sport RA.

\appendix

\section{VEVs for the Minkowski vacuum}

\label{sec:Mink}

In this section we consider the VEVs for the inertial vacuum in the $(D+1)$%
-dimensional locally Minkowski spacetime with spatial topology $%
R^{p+1}\times T^{q}$. Let us denote by $x_{\mathrm{M}}^{i}$ the
corresponding Cartesian coordinates with the line element $ds^{2}=\eta
_{il}dx_{\mathrm{M}}^{i}dx_{\mathrm{M}}^{l}$, where $\eta _{il}=\mathrm{diag}%
(1,-1,\ldots ,-1)$. The subspace covered by the coordinates $x_{\mathrm{M}%
}^{l}$, with $l=2,\ldots ,D$, is the same as in the Rindler problem, $x_{%
\mathrm{M}}^{l}=x^{l}$, $l=2,\ldots ,D$. The mode functions are specified by
the momentum $\mathbf{k}_{\mathrm{M}}=(k_{\mathrm{M}}^{1},\mathbf{k})$ where 
$\mathbf{k}$ coincides with the momentum variable for the Rindler modes (\ref%
{Modes}) and $-\infty <k_{\mathrm{M}}^{1}<+\infty $. The normalized mode
functions realizing the Minkowski vacuum become 
\begin{equation}
\varphi _{\mathrm{(M)}\mathbf{k}}^{(\pm )}(x)=\frac{2^{-\frac{p}{2}-1}e^{i%
\mathbf{k}_{\mathrm{M}}\cdot \mathbf{x}_{\mathrm{M}}\mp i\omega _{\mathrm{M}%
}t_{\mathrm{M}}}}{\pi ^{\frac{p+1}{2}}\sqrt{V_{q}\varepsilon _{\mathbf{k}}}}%
,\;\omega _{\mathrm{M}}=\sqrt{\left( k_{\mathrm{M}}^{1}\right)
^{2}+k^{2}+m^{2}}.  \label{Minkmodes}
\end{equation}%
The mode sums for the VEVs of the field squared and energy-momentum tensor
are expressed as%
\begin{eqnarray}
\left\langle \varphi \varphi ^{\dagger }\right\rangle _{\mathrm{(M)}} &=&%
\frac{1}{2V_{q}}\int \frac{d\mathbf{k}_{p+1}}{\left( 2\pi \right) ^{p+1}}%
\sum_{\mathbf{n}_{q}}\frac{1}{\omega _{\mathrm{M}}},  \notag \\
\langle T^{il}\rangle _{\mathrm{(M)}} &=&\frac{1}{2V_{q}}\int \frac{d\mathbf{%
k}_{p+1}}{\left( 2\pi \right) ^{p+1}}\sum_{\mathbf{n}_{q}}\frac{k^{i}k^{l}}{%
\omega _{\mathrm{M}}},  \label{TilM}
\end{eqnarray}%
where $i,l=0,1,\ldots ,D$, $\mathbf{k}_{p+1}=(k_{\mathrm{M}}^{1},\mathbf{k}%
_{p})$, and $k^{0}=\omega _{\mathrm{M}}$. Note that $\langle T^{il}\rangle _{%
\mathrm{(M)}}=0$ for $i,l=0,1,\ldots ,p+1$ and $i\neq l$. These VEVs are
given in the coordinate system $x_{\mathrm{M}}^{i}$. In order to obtain the
corresponding VEVs in the Rindler coordinates a coordinate transformation%
\begin{equation}
x_{\mathrm{M}}^{0}=t_{\mathrm{M}}=\xi \sinh \tau ,\;x_{\mathrm{M}}^{1}=\xi
\cosh \tau ,  \label{CoordTrans}
\end{equation}%
and $x_{\mathrm{M}}^{l}=x^{l}$ for $l=2,\ldots ,D$ is required. From the
Lorentz invariance in the uncompact subspace it follows that (no summation
over $l$) $\langle T_{l}^{l}\rangle _{\mathrm{(M)}}=\langle T_{0}^{0}\rangle
_{\mathrm{(M)}}$ for $l=1,2,\ldots ,p+1$. Now, it is easy to show that the
VEVs of the components $T_{l}^{i}$ coincide in the coordinate systems $x_{%
\mathrm{M}}^{i}$ and $x^{i}$.

In order to obtain the renormalized VEVs corresponding to (\ref{TilM}) we
will follow the zeta function regularization technique. The zeta function
related to the mode sums in (\ref{TilM}) is defined as%
\begin{equation}
\zeta (s)=\frac{1}{V_{q}}\int \frac{d\mathbf{k}_{p+1}}{\left( 2\pi \right)
^{p+1}}\sum_{\mathbf{n}_{q}}\frac{1}{\omega _{\mathrm{M}}^{2s}}.
\label{zeta}
\end{equation}%
In the special case $\tilde{\alpha}_{l}=0$ for all $l=p+2,\ldots ,D$, the
term with $\mathbf{n}_{q}=\mathbf{0}_{q}$ should be excluded from the sum.
For the analytic continuation of the zeta function to the physical points
for complex variable $s$, we first integrate over the momentum $\mathbf{k}%
_{p+1}$\ and then apply the generalized Chowla-Selberg formula \cite%
{Eliz98,Eliz12} to the series over $\mathbf{n}_{q}$. For the topological
contribution $\zeta _{\mathrm{t}}(s)$ this gives 
\begin{equation}
\zeta _{\mathrm{t}}(s)=\frac{2^{1-s}m^{D-2s}}{(2\pi )^{\frac{D}{2}}\Gamma (s)%
}\sideset{}{'}{\sum}_{\mathbf{n}_{q}}\cos (\mathbf{n}_{q}\cdot \boldsymbol{%
\tilde{\alpha}}_{q})f_{\frac{D}{2}-s}(mg_{\mathbf{n}_{q}}).  \label{zetatop}
\end{equation}

The renormalized VEVs of the field squared $\left\langle \varphi \varphi
^{\dagger }\right\rangle _{\mathrm{M}}$ and energy density $\langle
T_{0}^{0}\rangle _{\mathrm{M}}$ are directly obtained from here as $%
\left\langle \varphi \varphi ^{\dagger }\right\rangle _{\mathrm{M}}=\zeta _{%
\mathrm{t}}(1/2)/2$ and $\langle T_{0}^{0}\rangle _{\mathrm{M}}=\zeta _{%
\mathrm{t}}(-1/2)/2$. From the symmetry of the problem it follows that (no
summation over $i$) $\langle T_{i}^{i}\rangle _{\mathrm{M}}=\langle
T_{0}^{0}\rangle _{\mathrm{M}}$, $i=1,\ldots ,p+1$. In order to find the
components along compact dimensions we use the relation%
\begin{equation}
\frac{k^{i}k^{l}}{\omega _{\mathrm{M}}}=\frac{L_{i}L_{l}}{3}\partial _{%
\tilde{\alpha}_{i}}\partial _{\tilde{\alpha}_{l}}\omega _{\mathrm{M}%
}^{3}-\delta ^{il}\omega _{\mathrm{M}},  \label{Relder}
\end{equation}%
with $i,l=p+2,\ldots ,D$. The renormalized VEV is presented as%
\begin{equation}
\langle T^{il}\rangle _{\mathrm{M}}=\frac{1}{2}\left[ \frac{L_{i}L_{l}}{3}%
\partial _{\tilde{\alpha}_{i}}\partial _{\tilde{\alpha}_{l}}\zeta _{\mathrm{t%
}}(-3/2)-\delta ^{il}\zeta _{\mathrm{t}}(-1/2)\right] .  \label{TMink}
\end{equation}%
By using (\ref{zetatop}), one gets the expressions (\ref{phi2M}), (\ref{TiiM}%
), and (\ref{TilM2}).

\section{Alternative representations for the expectation values}

\label{sec:Alt}

In order to obtain alternative representations for the VEVs of the field
squared and energy-momentum tensor we use the formula \cite{Beze13}%
\begin{equation}
\sum_{\mathbf{n}}\cos (\mathbf{n}\cdot \boldsymbol{\beta })f_{\nu }(c\sqrt{%
b^{2}+\sum\nolimits_{i=1}^{r}a_{i}^{2}n_{i}^{2}})=\frac{(2\pi )^{r/2}}{%
a_{1}\cdots a_{r}c^{2\nu }}\sum_{\mathbf{n}}w_{\mathbf{n}}^{2\nu -r}f_{\nu -%
\frac{r}{2}}(bw_{\mathbf{n}}),  \label{Rel4}
\end{equation}%
where $\mathbf{n}=(n_{1},\ldots ,n_{r})$, $\boldsymbol{\beta }=(\beta
_{1},\ldots ,\beta _{r})$, and $w_{\mathbf{n}}^{2}=\sum\nolimits_{i=1}^{r}(2%
\pi n_{i}+\beta _{i})^{2}/a_{i}^{2}+c^{2}$. Taking $\beta _{i}=\tilde{\alpha}%
_{p+i+1}$, $a_{i}=L_{p+i+1}$, and 
\begin{equation*}
\nu =\frac{D-1}{2}+n,\;r=q=D-p-1,\;c=m,\;w_{\mathbf{n}}=\omega _{\mathbf{n}%
_{q}},
\end{equation*}%
we get%
\begin{equation}
\sum_{\mathbf{n}}\omega _{\mathbf{n}_{q}}^{p+2n}f_{\frac{p}{2}+n}(b\omega _{%
\mathbf{n}_{q}})=\frac{V_{q}m^{D-1+2n}}{(2\pi )^{\left( D-p-1\right) /2}}%
\sum_{\mathbf{n}_{q}}\cos (\mathbf{n}_{q}\cdot \boldsymbol{\tilde{\alpha}}%
)f_{\frac{D-1}{2}+n}(m\sqrt{b^{2}+\sum\nolimits_{i=p+2}^{D}L_{i}^{2}n_{i}^{2}%
}).  \label{SumTrans}
\end{equation}%
The application of this formula with $n=0$ and $b=2\xi \cosh y$ to the
series $\sum_{\mathbf{n}_{q}}\omega _{\mathbf{n}_{q}}^{p}f_{p/2}\left(
b\omega _{\mathbf{n}_{q}}\right) $ in (\ref{phi2}) leads to the
representation (\ref{phi22}).

The representations for the components $\langle T_{i}^{i}\rangle $ with $%
i=0,1,\ldots ,p+1$ are directly obtained from (\ref{Tii}) and (\ref{Fr})
applying to the series over $\mathbf{n}_{q}$ the formula (\ref{SumTrans})
with $n=1,2$. Special consideration is required for the part with $k_{r}^{2}$
in the expression for the stress $\langle T_{r}^{r}\rangle $ along the
compact dimension $x^{r}$, $r=p+2,\ldots ,D$, and for the off-diagonal
component. For the transformation of the term with $k_{r}^{2}$, we use the
relation 
\begin{equation}
\left[ u^{2\nu }f_{\nu }(u)\right] ^{\prime }=-u^{2\nu -1}f_{\nu -1}(u).
\label{relf2}
\end{equation}%
By using this relation, the following identity can be checked%
\begin{equation}
\sum_{\mathbf{n}_{q}}k_{r}^{2}\omega _{\mathbf{n}_{q}}^{p}f_{p/2}\left(
u\right) =L_{r}^{2}\frac{\partial ^{2}}{\partial \tilde{\alpha}_{r}^{2}}%
\sum_{\mathbf{n}_{q}}\omega _{\mathbf{n}_{q}}^{p+4}f_{p/2+2}(b\omega _{%
\mathbf{n}_{q}})+\sum_{\mathbf{n}_{q}}\omega _{\mathbf{n}%
_{q}}^{p+2}f_{p/2+1}(b\omega _{\mathbf{n}_{q}}).  \label{Ident}
\end{equation}%
For the series in the right-hand side we use (\ref{SumTrans}) with $n=1$ and 
$n=2$. This gives%
\begin{equation}
\sum_{\mathbf{n}_{q}}k_{r}^{2}\omega _{\mathbf{n}_{q}}^{p}f_{p/2}\left(
u\right) =\frac{V_{q}m^{D+1}}{(2\pi )^{\left( D-p-1\right) /2}}\sum_{\mathbf{%
n}_{q}}\cos (\mathbf{n}_{q}\cdot \boldsymbol{\tilde{\alpha}})\left[ f_{\frac{%
D+1}{2}}(w)-m^{2}L_{r}^{2}n_{r}^{2}f_{\frac{D+3}{2}}(w)\right] .
\label{rel2}
\end{equation}%
By using this formula for the corresponding part in $\langle
T_{r}^{r}\rangle $ and the formula (\ref{SumTrans}) for the remaining terms,
the representation given in (\ref{Zr}) is obtained.

For the transformation of the off-diagonal components (\ref{Til}) we use the
relation%
\begin{equation}
\frac{k^{i}k^{l}}{L_{i}L_{l}}\omega _{\mathbf{n}_{q}}^{p}f_{p/2}(b\omega _{%
\mathbf{n}_{q}})=\frac{\partial ^{2}}{\partial \tilde{\alpha}_{i}\partial 
\tilde{\alpha}_{l}}\left[ \omega _{\mathbf{n}_{q}}^{p+4}f_{p/2+2}(b\omega _{%
\mathbf{n}_{q}})\right] ,  \label{Ident2}
\end{equation}%
for $i\neq l$, $i.l=p+2,\ldots ,D$. This gives%
\begin{equation}
\langle T_{il}\rangle =\langle T_{il}\rangle _{\mathrm{M}}-\frac{L_{i}L_{l}}{%
\left( 2\pi \right) ^{p/2+1}V_{q}}\frac{\partial ^{2}}{\partial \tilde{\alpha%
}_{i}\partial \tilde{\alpha}_{l}}\int_{0}^{\infty }\frac{dy}{y^{2}+\pi ^{2}/4%
}\sum_{\mathbf{n}_{q}}\omega _{\mathbf{n}_{q}}^{p+4}f_{p/2+2}(b\omega _{%
\mathbf{n}_{q}}).  \label{Til2}
\end{equation}%
The series in this formula is transformed by the formula (\ref{SumTrans})
with $n=2$.

\end{document}